%% file: main.tex
\documentclass[conference]{IEEEtran}
\usepackage[nospace]{cite}
\usepackage{amsmath,amssymb,amsfonts}
\usepackage{graphicx}
\usepackage{textcomp}
\usepackage{xcolor}
\usepackage{fancyhdr}
\usepackage[hyphens]{url}

\usepackage[normalem]{ulem}
\usepackage{braket}

\usepackage[tight,footnotesize]{subfigure}
\usepackage{algorithm}
\usepackage{algpseudocode}
\usepackage{multirow}
\usepackage{makecell}

\usepackage{authblk}

\newcommand{\SQS}{QGo}
\newcommand{\tket}{\emph{t$|$ket$\rangle$}}

\long\def\comment#1{}

\def\BibTeX{{\rm B\kern-.05em{\sc i\kern-.025em b}\kern-.08em
    T\kern-.1667em\lower.7ex\hbox{E}\kern-.125emX}}

\title{QGo: Scalable Quantum Circuit Optimization \\Using Automated Synthesis} 

\author[1,*]{Xin-Chuan Wu\thanks{* Corresponding author: xinchuan@uchicago.edu}}

\author[2]{Marc Grau Davis}
\author[1]{Frederic T. Chong}
\author[3]{Costin Iancu}

\affil[1]{Department of Computer Science, University of Chicago, Chicago, IL 60637, USA}
\affil[2]{Electrical Engineering and Computer Science, MIT, Cambridge, MA 02139, USA}
\affil[3]{Computational Research Division, Lawrence Berkeley National Laboratory, Berkeley, CA 94720, USA}

\begin{document}
\maketitle
\pagestyle{plain}


\begin{abstract}
The current phase of quantum computing is in the Noisy Intermediate-Scale Quantum (NISQ) era. On NISQ devices, two-qubit gates such as CNOTs are much noisier than single-qubit gates, so it is essential to minimize their count. Quantum circuit synthesis is a process of decomposing an arbitrary unitary into a sequence of quantum gates, and can be used as an optimization tool to produce shorter circuits to improve  overall circuit fidelity. However, the time-to-solution of synthesis grows exponentially with the number of qubits. As a result, synthesis is intractable for circuits on a large qubit scale.

In this paper, we propose a  hierarchical, block-by-block optimization framework, \SQS, for quantum circuit optimization. Our approach allows an exponential cost optimization to scale to large circuits. \SQS{} uses a combination of partitioning and synthesis: 1) partition the circuit into a sequence of independent circuit blocks; 2) re-generate and optimize each block using quantum synthesis; and 3) re-compose the final circuit by stitching all the blocks together. We perform our analysis and show the fidelity improvements in three different regimes: small-size circuits on real devices, medium-size circuits on noise simulations, and large-size circuits on analytical models. Using a set of NISQ benchmarks, we show that \SQS{} can reduce the number of CNOT gates by 29.9\% on average and up to 50\% when compared with industrial compilers such as \tket. When executed on the IBM Athens system, shorter depth leads to higher circuit fidelity. We also demonstrate the scalability of our \SQS{} technique to optimize circuits of 60+ qubits. Our technique is the first demonstration of successfully employing and scaling synthesis in the compilation toolchain for large circuits.  Overall, our approach is robust for direct incorporation in production compiler toolchains.
\end{abstract}

\input{T_Introduction}

\input{T_Background}
\input{T_SQS}
\input{T_ExperimentalSetup}

\input{T_Evaluation}
\input{T_Extension}

\input{T_RelatedWork}
\input{T_Conclusion}

\input{T_Ack}

\bibliographystyle{IEEEtranS}
\bibliography{refs}

\end{document}

%% file: T_Introduction.tex
\section{Introduction}\label{sec:intro}
Quantum Computing (QC) is expected to solve certain computational problems that even the largest classical supercomputers cannot solve. QC algorithms might have significant impact on areas such as quantum chemistry \cite{peruzzo2014variational, kandala2017hardware}, cryptography \cite{shor1999polynomial}, and machine learning \cite{biamonte2017quantum}. Recently, QC devices up to 72 quantum bits (qubits) have been demonstrated by IBM and Google \cite{IBM, Google}. The current phase of QC is in the Noisy Intermediate-Scale Quantum (NISQ) era \cite{preskill2018quantum}. On NISQ devices, quantum gates are noisy and their count must be minimized to produce low-error circuits.  In particular, two-qubit gates such as CNOTs are much noisier than single-qubit gates. For NISQ devices, shorter depth translates directly into higher circuit fidelity.

Quantum circuit synthesis provides an orthogonal circuit optimization method, which generates a circuit from its high level mathematical description such as unitary matrix. There are several existing studies in this field \cite{shende2006synthesis,dawson2005solovay,di2016parallelizing, nagy2006implementation,al2015quantum,de2016block,davis2019heuristics,amy2013meet,kliuchnikov2012fast,martinez2016compiling,giles2013exact,tucci2005introduction,younis2020qfast,bocharov2015efficient}.  Given an arbitrary quantum circuit, we can compute the corresponding unitary, and then use synthesis to generate a new circuit, which has a different sequence of gates from the original circuit, but will be equivalent in terms of unitary operator performed.

However, synthesis algorithms face an ``exponential'' scalability challenge. For a $n$-qubit circuit, the unitary size is $2^n \times 2^n$. The solving time of synthesis scales exponentially with the number of qubits \cite{al2015quantum,di2016parallelizing, davis2019heuristics}. Thus, synthesis is intractable for circuits beyond a handful of qubits.

\begin{figure*}[!t] 

\subfigure[Original circuit]
{
\includegraphics[width=0.44\textwidth,keepaspectratio]{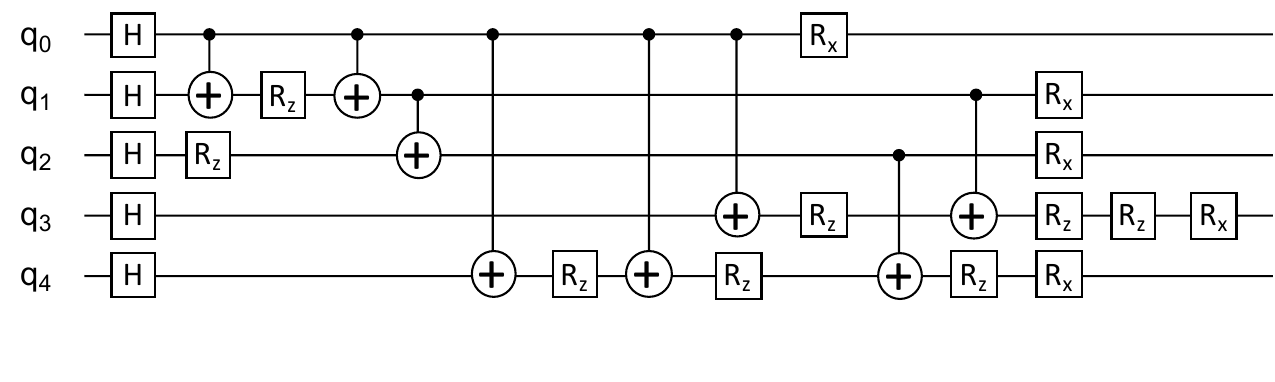}
\label{fig:example_ori}
}
\vspace{+3mm}
\subfigure[Partitioned circuit]
{
\includegraphics[width=0.54\textwidth,keepaspectratio]{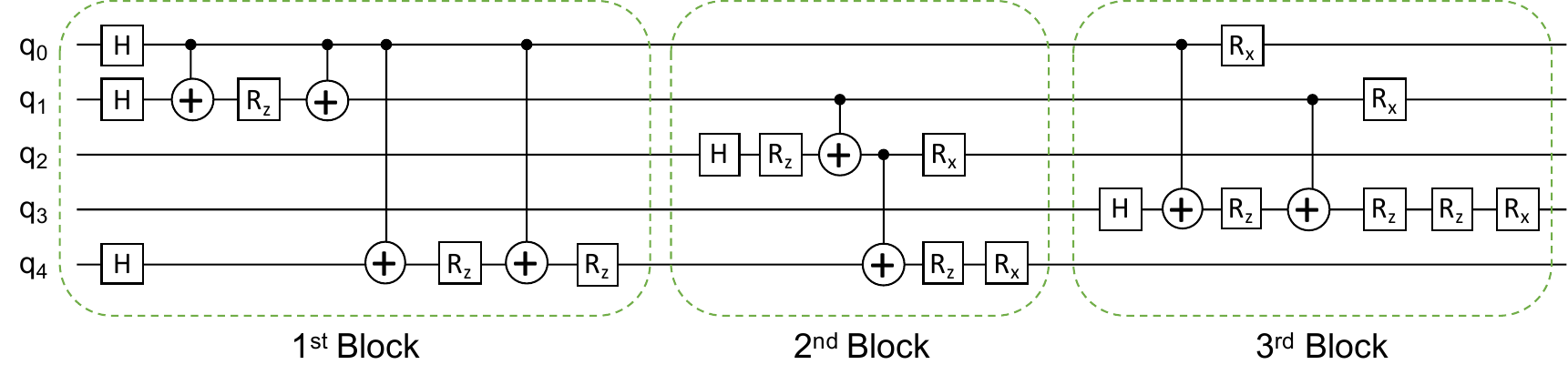}
\label{fig:example_blocks}
}
\vspace{+3mm}
\subfigure[Synthesized circuit]
{
\includegraphics[width=0.48\textwidth,keepaspectratio]{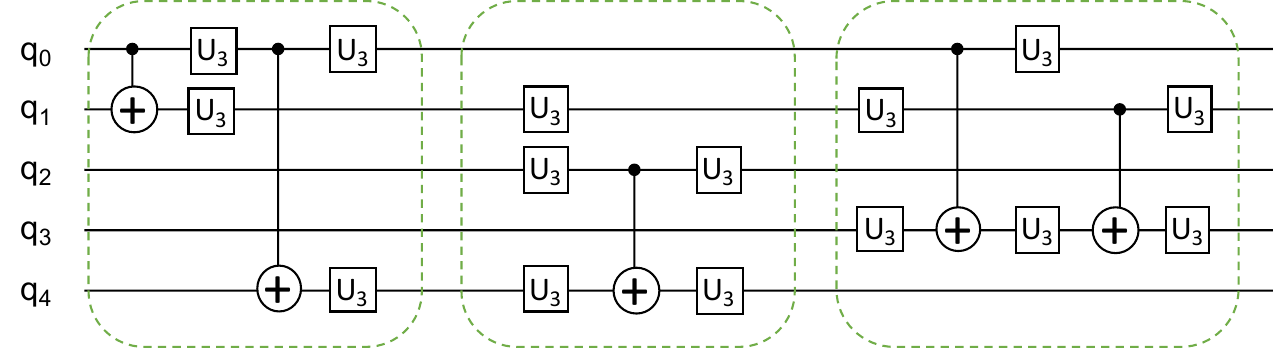}
}
\hspace{13mm}
\subfigure[Optimized circuit]
{
\includegraphics[width=0.3\textwidth,keepaspectratio]{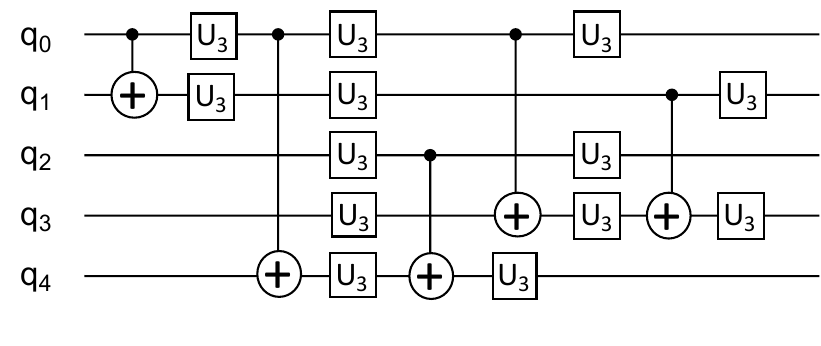}
}

\caption{An example of \SQS{} circuit optimization. (a) The original 5-qubit circuit. (b)The original circuit is partitioned into three blocks, and each block only contains gates on 3 qubits. The first circuit block is only on $q_0$, $q_1$, and $q_4$. The second block is on $q_1$, $q_2$, and $q_4$. The third block is on $q_0$, $q_1$, and $q_3$. All gates after partitioning still respect the gate dependency in the original circuit. (c) The synthesized circuit. After running quantum synthesis for each block, the CNOT counts in the first and second block are reduced, and the third block still has 2 CNOTs. (d) The single-qubit gates can be combined to produce the final optimized circuit. This circuit is effectively equivalent to the original circuit but with fewer CNOT gates.}
\label{fig:example}
\end{figure*}

To perform quantum synthesis for optimizing large scale circuits, we propose our hierarchical, block-by-block, optimization framework, called \SQS. \SQS{} uses a combination of partitioning and synthesis: 1) partition the circuit into independent sub-blocks whose size can be successfully handled by synthesis; 2) re-generate each block using synthesis; and 3)  re-assemble the circuit. Figure~\ref{fig:example} shows an example of \SQS{} circuit optimization that partitions a 5-qubit circuit into 3-qubit blocks. In general, partitioning a $n$-qubit circuit into multiple $k$-qubit blocks ($k<n$) leads to replacing an algorithm with $O(exp(n))$ complexity with a sequence of calls to $O(exp(k))$ algorithms. The time-to-solution grows linearly with the number of $k$-qubit blocks, which is a useful tradeoff when compared to the exponential scaling with the number of qubits. After all blocks are synthesized, \SQS{} then puts the blocks back together to produce the final optimized circuit.

\begin{figure}[!t]
\centering
\includegraphics[width=0.5\textwidth,keepaspectratio]{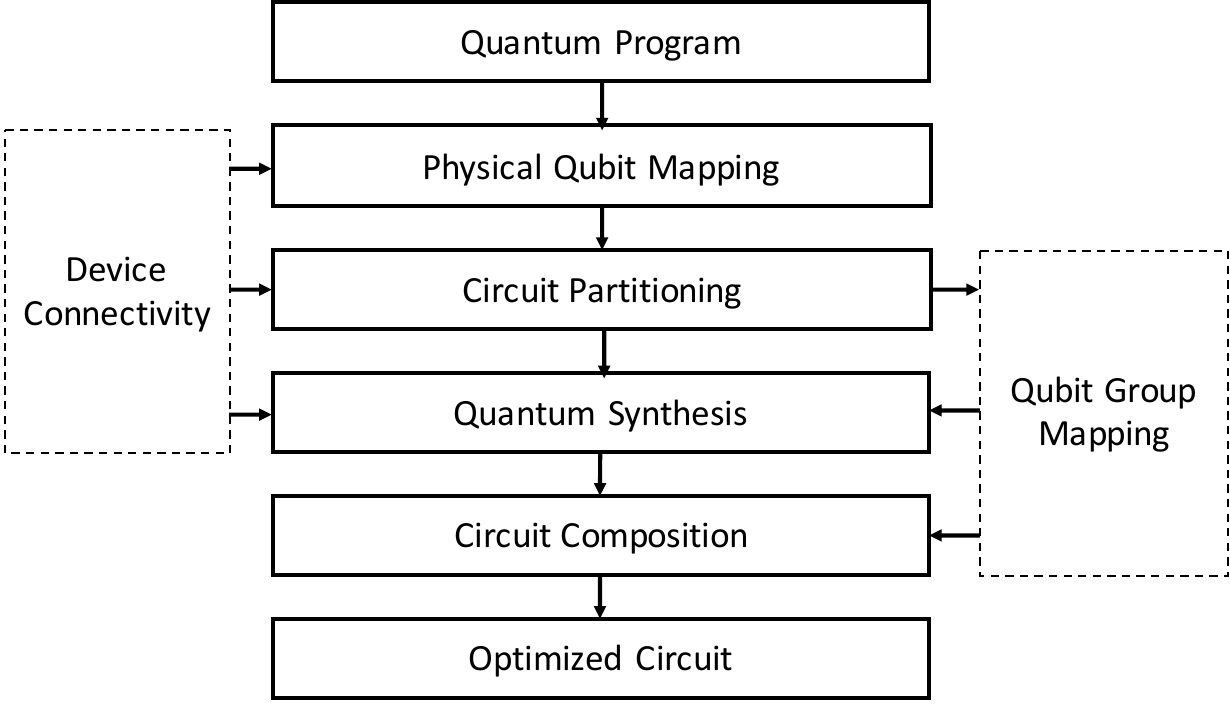}
\caption{Our compilation framework for scalable circuit optimization using synthesis (\SQS).}
\label{fig:toolflow}
\end{figure}


\SQS{} allows us to compile a quantum circuit to an optimized executable circuit for a target hardware. \SQS{} is a topology-aware compilation framework. The overall flow is described in Figure~\ref{fig:toolflow}. \SQS{} includes four core modules: physical qubit mapping, circuit partitioning, quantum synthesis, and circuit composition. The input to \SQS{} is a quantum circuit in a high-level quantum program, together with a description of the topology of the target processor. The first step in \SQS{} is topology mapping of the circuit to the target hardware. We leverage third party mappers and call into ``traditional'' compilers to perform mapping. This approach allows us to select the best mapper available for a given platform. We use the \tket{} compiler for physical qubit mapping as it is reported to effectively produce short circuits for several applications \cite{sivarajah2020t,cowtan2019phase}. We then run our partitioning algorithm to select tunable size blocks of the circuit that are independent of each other. In the synthesis procedure, each block is converted from its circuit format into a unitary matrix and re-synthesized with a synthesis tool. In this work, we use the state-of-the-art synthesis tool proposed in ~\cite{davis2019heuristics}. The final output of \SQS{} is the optimized circuit by stitching all the optimized blocks together.

For evaluation we use a series of NISQ circuits as our benchmarks. To understand the impact of the partitioning strategy we conduct experiments with 3- and 4-qubit blocks. We evaluate the quality of the generated circuits on IBM's Athens~\cite{IBMQX} device, as well as using noise simulation for medium-size circuits and on our analytical model for large-size circuits.

The  main  contributions  of  this  work  are:

\begin{itemize}
\item Our technique allows an exponential cost optimization tool to scale to large circuits, and is the first demonstration of successfully employing and scaling synthesis in the compilation toolchain for large circuit optimization. 


\item \SQS{} reduces the CNOT gate count well beyond the ability of existing compilers. Using a set of NISQ benchmarks, we show that \SQS{} can reduce the number of CNOT gates by 29.9\% on average and up to 50\% compared with circuits optimized by \tket{} compiler. These translate into direct fidelity improvements when  running on the IBM's Athens device~\cite{cross2018ibm}.

\item We evaluate fidelity improvements using 3 metrics for 3 scaling regimes: small-size circuits using fidelity on real devices, medium-size circuits using fidelity using simulations with noise, and large-size circuits using CNOT reduction measured statically by our compiler. To validate the trends shown by our simulations, we show that there is a 98\% correlation between our real-device results and our simulation results on small circuits.  To show that our CNOT reductions are also indicative of fidelity savings, we show there is a 78\% correlation to real devices on small circuits and 76\% correlation to simulation on medium circuits.

\item We present the sensitivity analysis by running Qiskit noise simulations to show the circuit fidelity improvements under different levels of gate errors. The results indicate that \SQS{} is important for NISQ devices.

\item We demonstrate the scalability of our technique to optimize the circuits of 60+ qubits. This demonstration suggests that \SQS{} provides a viable path towards higher fidelity for circuits on large scale qubits.
\end{itemize}

Our paper is organized as follows. In Section~\ref{sec:background}, we introduce the principles of quantum computation and quantum circuit synthesis techniques. In Section~\ref{sec:design}, we present our hierarchical, block-by-block, optimization tool flow and discuss in detail the techniques that make up our compilation framework. In Section~\ref{sec:setup}, we describe our benchmarks and experimental environment.  Section~\ref{sec:eval} evaluates the performance of our \SQS{} approach against industrial compilers. We discuss future directions in Section~\ref{sec:ext}. Section~\ref{sec:related} summarizes quantum circuit optimization studies. Finally, we provide conclusions in Section~\ref{sec:conclusion}.

%% file: T_Background.tex
\section{Background}\label{sec:background}

In this section, we briefly introduce the principles of quantum computation. We then present the relevant background on quantum circuit synthesis.

\subsection{Principles of Quantum Computation}
A qubit is a two-level quantum system, represented by two orthonormal computational basis states $\ket{0}$ and $\ket{1}$. The quantum state of a qubit can be described by any linear combination of $\ket{0}$ and $\ket{1}$: $\ket{\psi} = \alpha\ket{0} + \beta\ket{1}$, where $\alpha$ and $\beta$ are complex numbers satisfying $|\alpha|^2 + |\beta|^2 = 1$. More generally, the state of a $n$-qubit quantum system can be represented by using $2^n$ amplitudes.

A quantum circuit consists of a sequence of quantum gates on qubits, and a quantum gate is a unitary operator, $U$. The effect of a gate on a quantum state can be expressed as $\ket{\psi'} = U\ket{\psi}$. Gates are applied to one or many qubits simultaneously. Two-qubit controlled gates and arbitrary single-qubit gates are known to be universal \cite{divincenzo1995two}. CNOT gates and single-qubit rotation gates ($R_x, R_y, R_z$) constitute a commonly used universal gate set for quantum programming.


\subsection{Quantum Circuit Synthesis}
Quantum circuit synthesis is the process of taking a description of a desired unitary matrix and decomposing it into a sequence of smaller unitaries, representing the gates in a circuit that implements the target unitary.  One of the earliest techniques was the Solovay-Kitaev algorithm\cite{dawson2005solovay, nagy2006implementation, al2015quantum}, which combined a recursive matrix decomposition technique with numerical methods required for the base case.  This method served as a proof-of-concept for the field of synthesis, but the circuits it produces are very long.  One of the main goals of synthesis is to produce ``short'' circuits, minimizing metrics such as CNOT count, total gate count, and critical path length.  CNOT count is of particular importance on NISQ devices, since they contribute significantly more to the overall noise and runtime of circuits than single-qubit gates.  Because of this, CNOT count has been a particular target for synthesis approaches.  The approach shown in \cite{amy2013meet} is able to find minimal length circuits according to a weighting of different gates, and demonstrates a significant advantage over the Solovay-Kitaev approach.  The approaches shown in \cite{kliuchnikov2012fast} and \cite{di2016parallelizing} focus on synthesis runtime, which is another metric needed for synthesis to be practical.  All of these techniques mentioned so far work solely with discrete gatesets, where only a finite number of base gates are allowed.  However, most current quantum devices can perform continuously varying gates.  For example, superconducting qubits use virtual Z rotations, which can be performed with any angle, and trapped ion qubits can perform X and Y rotations with any angle.  These continuous gates can be used to perform any single qubit rotation, usually parameterized as $Z(\theta_1)X(\frac{\pi}{2})Z(\theta_2)X(\frac{\pi}{2})Z(\theta_3)$ or $X(\theta_1)Y(\theta_2)Z(\theta_3)$.  Leveraging this capability, the KAK decomposition can perform optimal 2-qubit synthesis to perform any 2 qubit unitary with at most 3 CNOTs \cite{tucci2005introduction}.  The approach shown in \cite{shende2006synthesis} uses the cosine-sine decomposition to produce universal gates for any number of qubits, which can perform any unitary simply by changing the parameters.  Other approaches use numerical optimization of parameterized gates to find high quality circuits \cite{martinez2016compiling, davis2019heuristics, younis2020qfast}.

We employ the search-based numerical synthesis method described in \cite{davis2019heuristics}.  This technique builds a search tree of CNOT structures filled in with parameterized single-qubit gates, and employs numerical optimization to evaluate the CNOT structure at each node in the search.  A* search is used to find the shortest CNOT structure that can implement the desired unitary, and numerical optimization is used to find parameters for the single-qubit gates to produce a fully instantiated circuit.  We choose this technique because of its ability to minimize CNOT count.



%% file: T_SQS.tex
\section{Scalable Circuit Optimization using Synthesis}\label{sec:design}
Quantum synthesis is a powerful tool for circuit optimization. However, it is not scalable for large circuits. To optimize asymptotically large circuits, we develop our block-by-block optimization technique, called \SQS. \SQS{} allows us to compile a high-level quantum program to a hardware topology-aware executable circuit. \SQS{} takes the device qubit connectivity and a quantum program as inputs, and produces a compiled executable circuit, minimizing the CNOT gate count by using synthesis. 

\subsection{\SQS{} Overview}
Figure~\ref{fig:toolflow} shows an overview of \SQS. It contains four core compiler components: physical qubit mapping, circuit partitioning, quantum synthesis, and the circuit composition procedure. The inputs for \SQS{} consist of a high-level quantum program and a target device connectivity description. First, \SQS{} performs physical qubit mapping to assign the logical qubits to physical qubits and to resolve all unexecutable two-qubit gates by adding swap gates. Since our goal is to produce an optimized topology-aware circuit, if we perform physical qubit mapping after quantum synthesis, there will be swap gates inevitably added into the final circuit. Therefore, we perform physical qubit mapping before quantum synthesis, so that the swap gates added into the circuit become a part of the input circuit for the circuit synthesis, and hence the final circuit would have a reduced CNOT gate count. Second, \SQS{} partitions the circuit into multiple small circuit blocks. Each block only interacts with a few qubits. The qubit group mapping is generated to indicate what qubits are involved for each circuit block. This qubit group mapping information will be used for the later circuit composition procedure. Since the number of qubits in a block reflects the size of the corresponding unitary matrix, we demonstrate the partitioning of 3-qubit and 4-qubit blocks in this work, but the block size can be further increased. In general, a larger unitary matrix would take exponentially longer time to decompose into a sequence of gates. A circuit partitioned with a larger block size would have fewer blocks, and each block tends to allow the synthesizer to decide circuit elements rather than the mapping algorithm, and hence the final optimized circuit would have fewer CNOTs. Thus, the block size is a trade-off between the time-to-solution and the quality of solution (CNOT reduction). Next, quantum synthesis is applied to the partitioned circuit blocks. We use the state-of-the-art synthesis tool proposed in \cite{davis2019heuristics} for our work. Finally, \SQS{} performs the circuit composition by following the qubit group mapping to concatenate the synthesized blocks to produce the final optimized circuit.

\subsection{Physical Qubit Mapping}\label{subsec:physical}
Given an input quantum circuit and the qubit connectivity graph, physical qubit mapping is to find an initial qubit mapping and insert swap gates to satisfy all two-qubit interactions and try to minimize the total number of swap gates and circuit depth in the final hardware-compliant circuit. Physical qubit mapping problem is NP-Complete \cite{siraichi2018qubit}. Several approaches for solving this problem have been proposed~\cite{bahreini2015minlp,wille2014optimal,lye2015determining,cross2018ibm,maslov2008quantum,chakrabarti2011linear,nishio2019extracting,li2019tackling, itoko2020optimization,sivarajah2020t,wu2020tilt, zulehner2018efficient}. 

We design the physical qubit mapping as the first step of \SQS{} because we want to have the additional swaps gates to be part of the input circuit for the remaining optimization process, so that the optimization by quantum synthesis can reduce the CNOT gate count in the final circuit. In addition, we also find that most quantum applications have certain repeated CNOT patterns, and existing qubit mapping algorithms will perform better by leveraging the structure of these patterns. If the synthesis process runs before physical qubit mapping, the  pattern will be destroyed, and the number of additional swap gates will increase in some cases. Thus, performing physical qubit mapping before synthesis can benefit the overall optimization.

We compare industrial compilers such as IBM Qiskit \cite{cross2018ibm} and \tket{} compilers \cite{sivarajah2020t}. Since the \tket{} compiler produces shorter circuits in our experiments and is reported to produce shorter circuits for several quantum applications compared with other techniques~\cite{cowtan2019phase}, we adopt the \tket{} compiler for physical qubit mapping in our \SQS.

\subsection{Circuit Partitioning}

\begin{table}[h!]
\centering
\caption{Notations used in our algorithm.}
\small
\begin{tabular}{ll}
\hline\hline
Notation & Definition\\
\hline
$n$ & The total number of qubits in a circuit\\
\hline
$k$ & The total number of qubits in a partitioned block\\
\hline
$g_i$ & A gate operation. $i$ is the index of the gate. \\
\hline
$q_i$ & A qubit. $i$ is the index of the qubit.\\
\hline
$Q_i$& A qubit group. $i$ is the index of the group.\\
\hline
$E_{Q_i}$ & A set of executable gates on $Q_i$\\
\hline
$G$ & Gate dependency graph\\
\hline
$B$ & A list of partitioned blocks\\
\hline
$M$ & A list of qubit group mapping\\
\hline\hline
\end{tabular}
\label{tab:notation}
\end{table}

\SQS{} partitions a circuit into multiple small circuit blocks. Each block contains gates only on a small group of qubits. Figure~\ref{fig:example} shows an example of partitioning. Each circuit block consists of 3 qubits (Figure~\ref{fig:example_blocks}). Since the CNOT reduction would be higher when there are more CNOTs in a block, the goal of partitioning is to \emph{find a block with the number of qubits and biggest number of CNOTs}. Thus, we propose our algorithm: we use a greedy-based heuristic approach. This is an efficient approach; compared with other partitioning algorithms such as dynamic programming that would be limited by circuit depth, our heuristic algorithm is scalable for large-scale circuit partitioning. We summarize the notations used in our algorithm in Table~\ref{tab:notation}.

Before our heuristic algorithm, we define a few terms used in our algorithm.

\textbf{Qubit group.} A qubit group is a set of qubits in a circuit block. For a $n$-qubit circuit, there are ${n \choose k}$ different combinations of qubit groups for a $k$-qubit block. For example, the circuit shown in Figure~\ref{fig:example_ori} is a 5-qubit circuit. For partitioning the circuit into 3-qubit blocks, $\{q_0, q_1, q_2\}, \{q_0, q_1, q_3\},...,$ and $\{q_2, q_3, q_4\}$ are possible qubit groups for a block. Figure~\ref{fig:example_blocks} shows the circuit after partitioning. The qubit group of the 1st block is $\{q_0, q_1, q_4\}$, and the 2nd block is $\{q_1, q_2, q_4\}$, and the 3rd block is $\{q_0, q_1, q_3\}$. We use $Q_i$ to denote a qubit group in this paper, where $i = 1,2,...,{n \choose k}$. The qubit group size $k$ is limited to a small number due to the limitation of quantum synthesis. In this paper, we focus our analysis on the size of 3 and 4 qubits in a qubit group.

\begin{figure}[!t]
\centering
\includegraphics[width=0.26\textwidth,keepaspectratio]{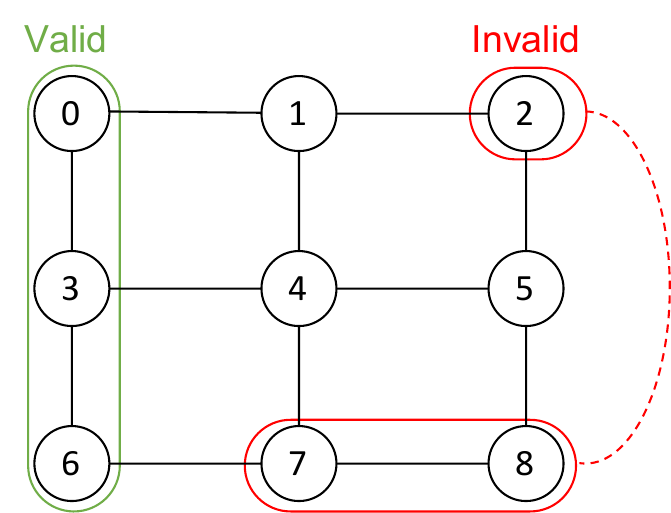}
\caption{Example of valid and invalid qubit groups.}
\label{fig:valid_qg}
\end{figure}

\textbf{Valid qubit group.} Since the input circuit for our partitioning algorithm is a physical qubit mapped circuit, all two-qubit gates are applied on the neighbor qubits. We can define \emph{valid qubit groups} by considering device connectivity to reduce the search space in our algorithm. We map the qubit group onto the device connectivity graph. If a qubit group is a connected component, it is a valid qubit group; otherwise, it is an invalid group. Figure~\ref{fig:valid_qg} shows an example of a valid qubit group. Considering a $3\times3$ grid connectivity, the qubit group $\{q_0, q_3, q_6\}$ is a valid qubit group. However, the qubit group $\{q_2, q_7, q_8\}$ is an invalid qubit group. When partitioning a circuit, we only need to consider valid qubit groups.

\begin{figure}[!t] 
\centering
\subfigure[Original circuit]
{
\includegraphics[width=0.42\textwidth,keepaspectratio]{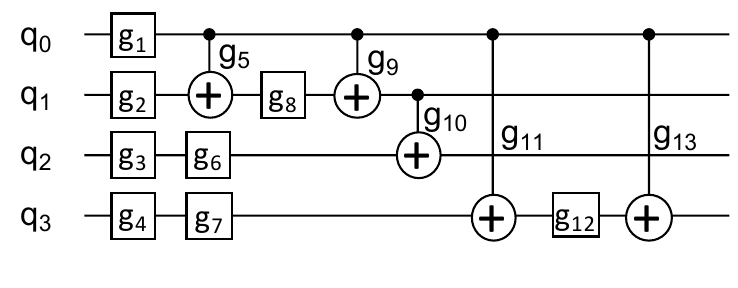}
\label{fig:dag_example_ori}
}
\vspace{4mm}
\subfigure[Gate dependency graph]
{
\includegraphics[width=0.42\textwidth,keepaspectratio]{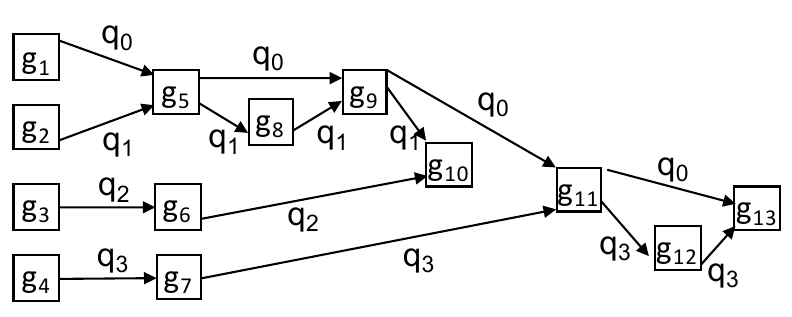}
\label{fig:dag_example}
}

\vspace{4mm}
\subfigure[Executable gates on qubit group $\{q_0,q_1,q_2\}$.]
{
\includegraphics[width=0.42\textwidth,keepaspectratio]{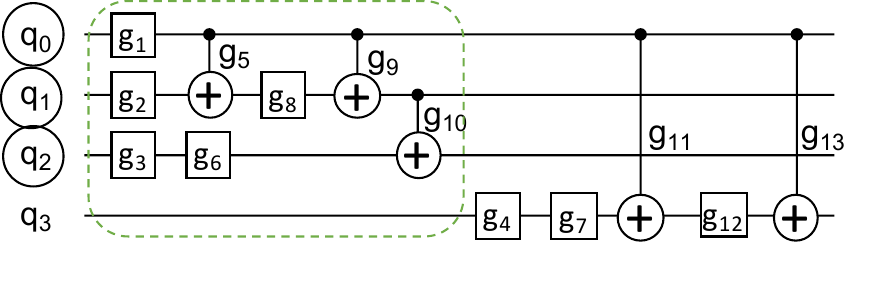}
\label{fig:qg_012}
}

\vspace{4mm}
\subfigure[Executable gates on qubit group $\{q_0,q_1,q_3\}$.]
{
\includegraphics[width=0.42\textwidth,keepaspectratio]{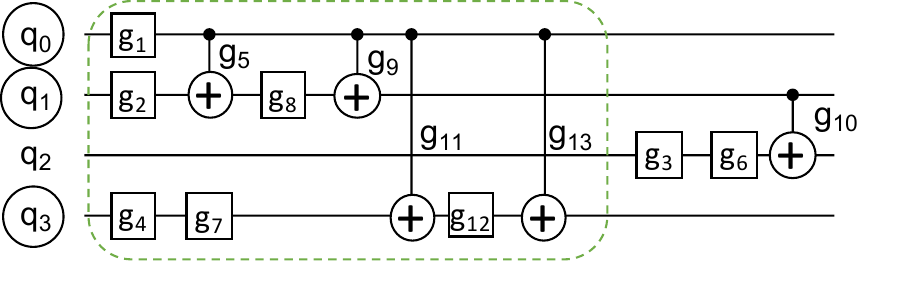}
\label{fig:qg_013}
}

\caption{An example of gate dependency graph and executable gates on a qubit group.}
\label{fig:gdg}
\end{figure}

\textbf{Gate dependency graph.} We use a Directed Acyclic Graph (DAG) to represent the gate dependency of a circuit. In a DAG, a node represents a gate, and an edge is the qubit involved for the gate. A gate can be executed only when all the previous gates have been executed. Figure~\ref{fig:gdg} shows an example of gate dependency graph. The DAG in Figure~\ref{fig:dag_example} is generated from the original circuit (Figure~\ref{fig:dag_example_ori}). For example, the gate $g_5$ depends on $g_1$ and $g_2$ because $q_0$ is used for $g_1$ and $g_5$, and $q_1$ is used for $g_2$ and $g_5$. Thus, $g_5$ can be executed only after both $g_1$ and $g_2$ are executed.

\textbf{Executable gates on a qubit group.} A single-qubit gate on $q_i$ can count as an executable gate on a qubit group $Q_i$ only when $q_i \in Q_i$ and all previous gates on $q_i$ are executable gates on $Q_i$. A two-qubit gate on ($q_i, q_j$) can count as an executable gate on a qubit group $Q_i$ only when both $q_i, q_j \in Q_i$ and all previous gates on $q_i$ and $q_j$ are executable gates on $Q_i$. In Figure~\ref{fig:gdg}, for example, when we count the executable gates on the qubit group $\{q_0,q_1,q_2\}$, the gate $g_9$ is an executable gate because $q_0$ and $q_1$ are in the qubit group and all the previous gates on $q_0$ and $q_1$ are also executable gates. However, $g_4$ is not an executable gate on this qubit group because $q_3$ is not in the target qubit group. We use $E_{Q_i}$ to denote the largest set of executable gates on the qubit group $Q_i$. We can get each $E_{Q_i}$ by traversing the gate dependency graph. Figure~\ref{fig:qg_012} and Figure~\ref{fig:qg_013} show $E_{\{q_0,q_1,q_2\}}$ and $E_{\{q_0,q_1,q_3\}}$, respectively. 

\begin{algorithm}[t]
 \caption{Circuit Partitioning} \label{algo:partition}
 \begin{algorithmic}[1]
 \State Convert the circuit into a gate dependency graph $G$.
 \State $B \gets \phi$
 \State $M \gets \phi$
 \While {$G$ is not empty}
  \For {each valid qubit group $Q_i$}
    \State $E_{Q_i} \gets \{\text{Executable Gates on }Q_i\}$
    \State Compute $Score(E_{Q_i})$
  \EndFor
  \State Select the $E_{Q_m}$ with the maximal score;
  \State $B.append(E_{Q_m})$
  \State $M.append(Q_m)$
  \State $G \gets G - E_{Q_m}$
 \EndWhile
 \State Output the sequence of circuit blocks $B$ and the list of qubit group mapping $M$
 \end{algorithmic}
\end{algorithm}

Algorithm~\ref{algo:partition} shows our partitioning procedure. First, we prepare the gate dependency graph $G$ for the target circuit. Next, we collect the set of executable gates, $E_{Q_i}$, for each qubit group, and also compute the score for each $E_{Q_i}$. Since the objective is to find the biggest circuit block, we select the $E_{Q_m}$ with the maximal score as the partitioned block, and save the $Q_m$ as the qubit group mapping for this block. Once a circuit block is partitioned, we remove the block from the gate dependency graph $G$, and repeat the procedure to find the next circuit block partition until the gate dependency graph $G$ is empty, which means the entire circuit is partitioned. Finally, a sequence of circuit blocks $B$ and a list of qubit group mapping $M$ are the output of the circuit partitioning process.

The heuristic cost function indicates the number of executable gates for a qubit group. The general form of our heuristic cost function for a qubit group $Q_i$ is shown as follows:
\begin{equation}
 Score(E_{Q_i}) = N_{E_{Q_i}},
\end{equation}
where $N_{E_{Q_i}}$ is the number of executable CNOT gates on $Q_i$. Since the objective of partitioning is to find a block for quantum synthesis to minimize the CNOT count, a block with more CNOT gates can potentially achieve higher CNOT reduction. Thus, we use CNOT count as the score function.

The time complexity of our heuristic algorithm for partitioning a circuit into $k$-qubit blocks is $O(n^kg)$, where $n$ is the number of qubits in the original circuit, $g$ is the total number of gates, $k$ is the number of qubits in a partitioned block.

\subsection{Quantum Circuit Synthesis}\label{subsec:synthesis}
After the circuit is partitioned into multiple blocks, \SQS{} applies quantum synthesis for each circuit block. We integrate the state-of-the-art synthesis tool proposed in \cite{davis2019heuristics} into our \SQS{} framework. For each circuit block, \SQS{} computes the corresponding unitary matrix, and this matrix is the input for the synthesis process. As the synthesized circuit should respect the hardware topology, the qubit group for each block and the device connectivity are also the input parameters for the synthesis tool. With our block-based synthesis scheme, the time-to-solution for a $n$-qubit circuit is reduced from $O(exp(n))$ to $O(exp(k))$. 

After the synthesis, there is an undesired synthesis distance due to numerical approximation, leading the final unitary at a distance from the original unitary~\cite{nielsen2002quantum,davis2019heuristics,younis2020qfast}.  As a result, the final unitary distance is in the range of $10^{-10}$ to $10^{-15}$. In the NISQ era, when running a quantum circuit on a real machine, the unitary executed is different from the original intended unitary due to the presence of noise. Since gate errors are multiple orders of magnitudes larger than synthesis distances, these synthesis distances are insignificant. In Section~\ref{subsec:infidelity} we will show that this distance only causes negligible impact on the overall state fidelity.

\subsection{Circuit Composition}
Once all circuit blocks have been synthesized, \SQS{} composes the entire circuit by stitching all circuit blocks. The list of qubit group mappings provide the qubit mapping information to connect blocks correctly. In general, the number of CNOTs in the synthesized block is less than in the original block. If the synthesized circuit block has an equal or greater number of CNOTs compared to the original block, \SQS{} will choose to use the original circuit block to the circuit composition to avoid unnecessary synthesis distance due to floating point errors. Once all blocks are put together, \SQS{} combines the adjacent single-qubit gates to further reduce the gate count, and produces the final optimized circuit.

%% file: T_ExperimentalSetup.tex
\section{Experimental Setup}\label{sec:setup}

In this section, we describe a set of benchmarks used in our evaluation as well as the experimental parameters.

\subsection{Benchmarks}
\begin{table}[h!]
\footnotesize
\centering
\caption{List of benchmarks.}
\begin{tabular}{cccl}
\hline\hline
Size & Application  & Qubits & Description\\
\hline
\multirow{6}{*}{Small} & QAOA5&	5 & QAOA on MaxCut problem \\
& TFIM5&	5 & Transverse-field Ising model \\
& MUL5&  5 & Multiplier arithmetic function \\
& ADDER4& 4 & Adder arithmetic function \\
& QFT5&	5 & Quantum Fourier transform \\
& HLF5&	5 & Hidden linear function \\
\hline
\multirow{6}{*}{Medium} & QAOA10&	10 & QAOA on MaxCut problem \\
& TFIM10&	10 & Transverse-field Ising model \\
& MUL10&  10 & Multiplier arithmetic function  \\
& ADDER9& 9 & Adder arithmetic function \\
& QFT9&	9 & Quantum Fourier transform \\
& HLF9&	9 & Hidden linear function \\
\hline
\multirow{3}{*}{Large} & MUL60&  60 & Multiplier arithmetic function \\
& ADDER63& 63 & Adder arithmetic function \\
& QFT64&	64 & Quantum Fourier transform \\
\hline\hline\\
\end{tabular}
\label{tab:apps}
\end{table}

To evaluate \SQS{} against real applications, we select multiple important quantum applications as our benchmarks. Table~\ref{tab:apps} lists the applications and brief description in our evaluation. We have three sets of applications according to the different sizes of the circuits. The first set consists of small-size circuits on 4 and 5 qubits; the second set is the medium-size of 9- and 10-qubit circuits; and the third set of benchmarks, large-size circuits, contains circuits beyond 60 qubits. In our evaluation, we run the small-size circuits on a 5-qubit quantum device, Athens, provided by IBM quantum experience platform to demonstrate how much circuit fidelity is improved through our optimization on a real NISQ machine. Next, we run our medium-size benchmarks by using Qiskit noisy simulators to show the fidelity improvement under different levels of gate errors. Finally, we use the large-size benchmarks to demonstrate the scalability of our \SQS{} technique.

Quantum Approximate Optimization Algorithm (QAOA) is a hybrid quantum-classical variational algorithm and it is one of the most important quantum algorithms in the NISQ era~\cite{farhi2014quantum}. In our study, the QAOA application is a hardware-efficiency ansatz~\cite{moll2018quantum} for MaxCut problem.  Transverse Field Ising Model (TFIM) is for problems that study the time evolution of chemical systems \cite{shin2018phonon, heyl2013dynamical}. Multiplier (MUL) \cite{gidneycirq} and adder \cite{gidneycirq,cuccaro2004new} benchmarks are important arithmetic functions in several quantum applications. Quantum Fourier transform (QFT) \cite{javadiabhari2015scaffcc} is a kernel function in many quantum algorithms such as phase estimation algorithm \cite{cleve1998quantum}, Shor's algorithm \cite{shor1999polynomial}, and the algorithm for hidden subgroup problem \cite{jozsa2001quantum}. Hidden Linear Function (HLF) is a quantum circuit solving a problem from a previous study~\cite{bravyi2018quantum}.

\begin{figure}[!t]
\centering
\includegraphics[width=0.34\textwidth,keepaspectratio]{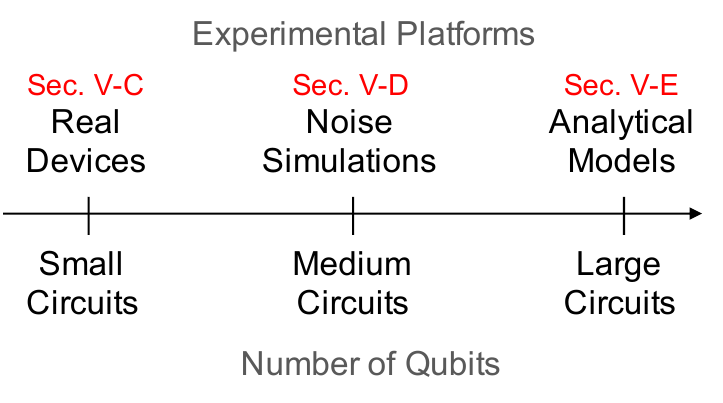}
\caption{Experimental platform used in our evaluation.}
\label{fig:exp}
\end{figure}

\subsection{Experimental Parameters}
We evaluate our \SQS{} with 3-qubit and 4-qubit blocks on different sizes of circuits. All \SQS{} processes and noisy simulations are carried out on a Ubuntu 16.04 system with Intel Xeon Silver 4110 32-core CPU (2.1 GHz) and 128GB RAM.

Figure~\ref{fig:exp} summarizes our fidelity analysis. For small circuits, we show the experimental results on IBM's Athens device (Section~\ref{subsec:real}). For medium circuits, we run the experiments on noise simulations (Section~\ref{subsec:sim}). Finally, we use our analytical model to perform the fidelity results for large circuits (Section~\ref{subsec:model}). The \tket{} compiler with the highest optimization level is the baseline in our evaluation. The \tket{} compiler is reported to effectively produce shorter circuits for several applications compared with other compilers~\cite{sivarajah2020t,cowtan2019phase}.

%% file: T_Evaluation.tex
\begin{figure*}[!t]
\centering
\subfigure[Total number of CNOTs in the circuits optimized by the baseline compiler and \SQS{}.]
{
\includegraphics[width=0.66\textwidth,keepaspectratio]{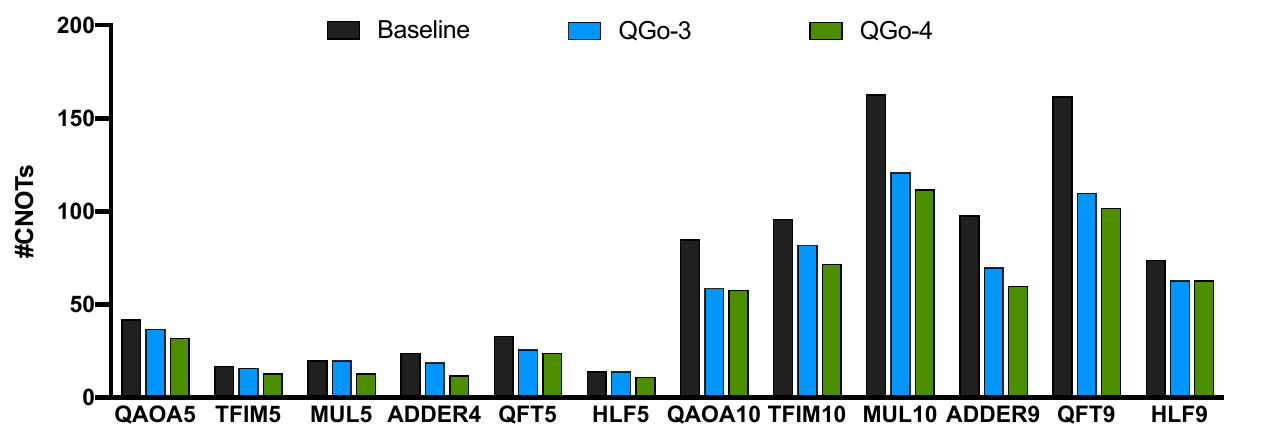}
\label{fig:num_CNOT_small}
}

\subfigure[CNOT reduction rate compared with the baseline compiler.]
{
\includegraphics[width=0.66\textwidth,keepaspectratio]{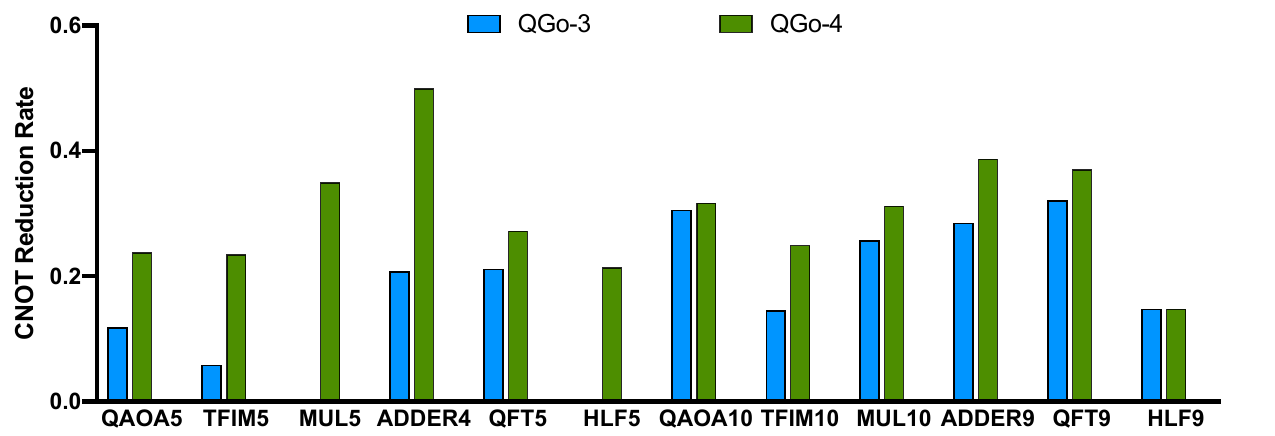}
\label{fig:CNOT_reduction_small}
}
\caption{The numbers of CNOTs are reduced in the \SQS{}-optimized circuits. In general, \SQS-4 can achieve higher CNOT reduction rate than \SQS-3.}
\label{fig:CNOT_small}
\end{figure*}

\begin{figure*}[!t]
\centering
\includegraphics[width=0.66\textwidth,keepaspectratio]{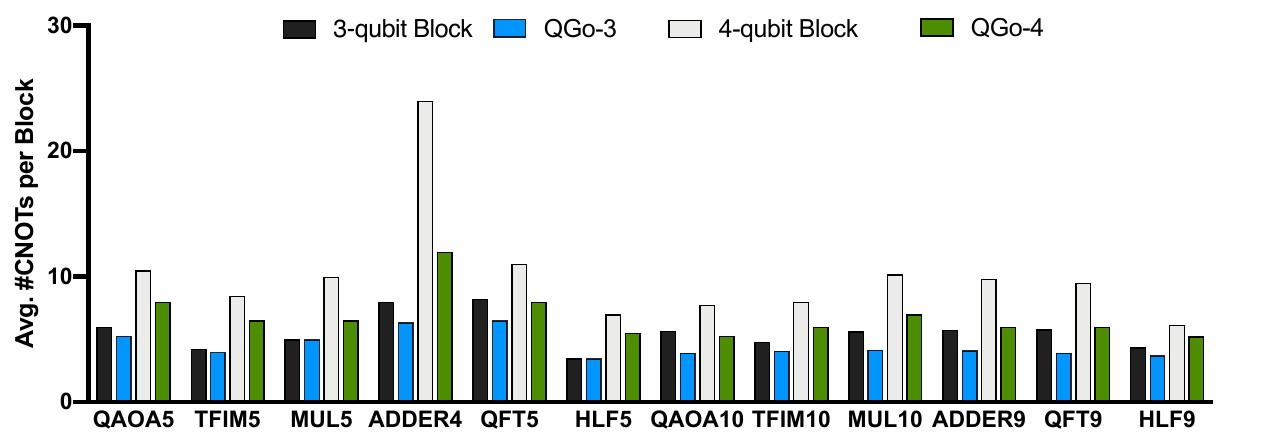}
\caption{Average CNOT count in a block.}
\label{fig:avg_CNOT_small}
\end{figure*}

\section{Evaluation}\label{sec:eval}

In our evaluation, we first show the CNOT reduction in the \SQS-optimized circuits compared to the baseline (\tket) optimization, and we compare the compilation time using 3-qubit and 4-qubit blocks. We then discuss the impact of synthesis distance on the quantum state. Next, we present the results running on a real quantum device, IBM's 5-qubit device (Athens), to prove that our technique is critical for the current devices. Then a sensitivity analysis is performed to show how much circuit fidelity is improved under different gate errors. We also demonstrate the scalability of \SQS{} by optimizing large circuits.

\subsection{CNOT Reduction}
\begin{figure}[!t]
\centering
\includegraphics[width=0.46\textwidth,keepaspectratio]{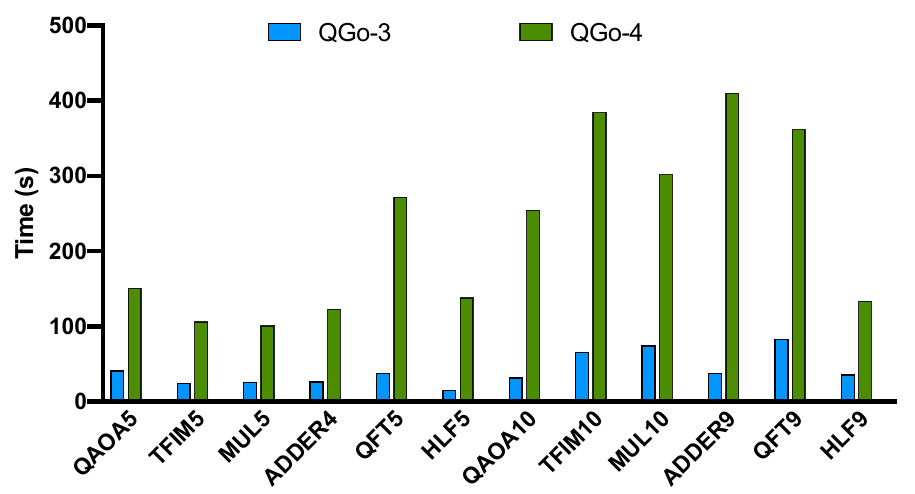}
\caption{Compilation time. A block with larger size takes more time to synthesize the circuit block. }
\label{fig:time_small}
\end{figure}

We use the \tket{} compiler to map 4- and 5-qubit circuits on the IBM's Athens topology, which is a linear connectivity, and map 9- and 10-qubit circuits on a 2D lattice of physical qubits. As discussed in Section~\ref{subsec:physical}, we compare industrial compilers such as IBM Qiskit \cite{cross2018ibm} and \tket{} compilers \cite{sivarajah2020t}. Since the \tket{} compiler is shown to generate shorter circuits for several quantum applications~\cite{cowtan2019phase}, we use the circuits optimized by \tket{} compiler as our baseline. The circuits optimized using our \SQS{} with 3-qubit blocks and 4-qubit blocks are denoted as \SQS-3 and \SQS-4, respectively. Figure~\ref{fig:CNOT_small} shows the total number of CNOTs in the circuits optimized by the baseline and \SQS. Our \SQS{} optimization reduces the total number of CNOTs. For MUL5 and HLF5, \SQS-3 does not reduce the CNOT count because there is only a small number of CNOTs in a block. \SQS-4 can reduce the CNOT counts across all benchmarks. The average CNOT reduction of \SQS-4 is 29.9\%. In general, \SQS-4 can achieve more CNOT reduction when compared with \SQS-3 because a large block will have more CNOTs in a block, and this is easier for the synthesis to find a circuit using less CNOTs than the original circuit block. Figure~\ref{fig:avg_CNOT_small} shows the average CNOT count in a block. If a circuit is partitioned into 4-qubit blocks, the average CNOT count per block is higher when compared with 3-qubit blocks. As a result, \SQS-4 can achieve higher CNOT reduction. However, the time-to-solution of \SQS-4 is much longer than \SQS-3 (Figure~\ref{fig:time_small}).

\subsection{Impact of Synthesis Distance} \label{subsec:infidelity}
\begin{figure}[!t]
\centering
\includegraphics[width=0.46\textwidth,keepaspectratio]{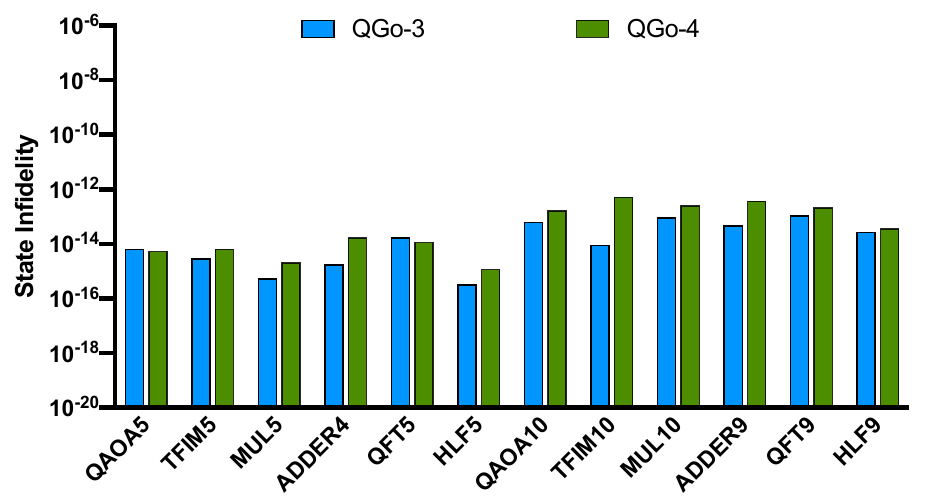}
\caption{State infidelity due to the synthesis distance. Compared with the gate error on NISQ devices, the state infidelity due to synthesis distance is negligible.}
\label{fig:state_infidelity_small}
\end{figure}

As discussed in Section~\ref{subsec:synthesis}, since there is a synthesis distance due to numerical approximation, the final state is not exactly the same as the ideal state of the original circuit. In order to analyze the impact of synthesis distance, we use ideal simulation to obtain the final states of the original circuit and synthesized circuit, and calculate the state infidelity \cite{nielsen2002quantum}. Figure~\ref{fig:state_infidelity_small} shows the state infidelity. Since the unitary matrix distance of 4-qubit circuit synthesis is fundamentally larger than 3-qubit circuit synthesis, \SQS-4 has slightly larger state infidelity when compared with \SQS-3. 
The state infidelity of each optimized circuit is less than $10^{-12}$.  This is insignificant when compared with the gate error in the NISQ era. The current gate infidelity on real devices is ranged from $10^{-1}$ to $10^{-6}$. Thus, the impact of synthesis distance is negligible.

\subsection{Running on Real Hardware}\label{subsec:real}
\begin{figure}[!t]
\centering
\includegraphics[width=0.42\textwidth,keepaspectratio]{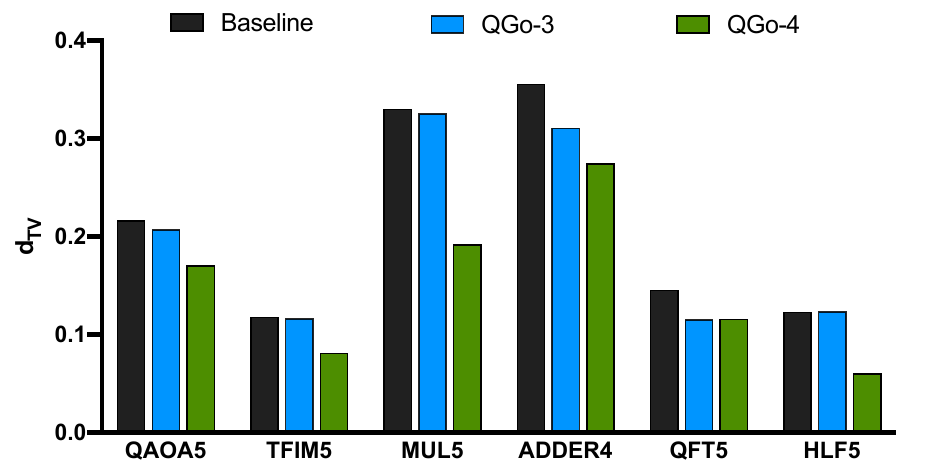}
\caption{Running optimized small circuits on IBM's Athens, a 5-qubit device. (Lower $d_{TV}$ is better.)}
\label{fig:tvd_q5}
\end{figure}

\begin{figure}[!t]
\centering
\includegraphics[width=0.36\textwidth,keepaspectratio]{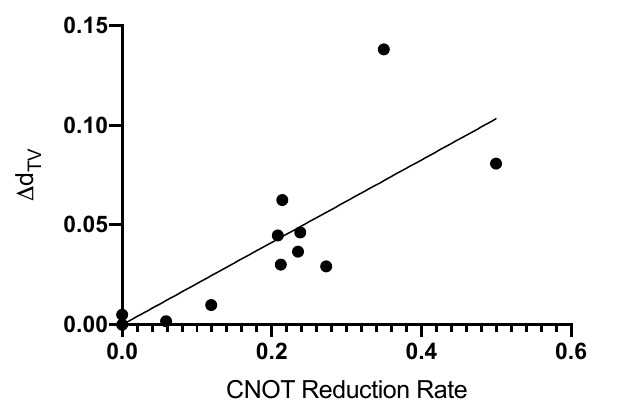}
\caption{Correlation between CNOT reduction rate and $d_{TV}$ improvement measured on IBM's Athens. Each data point is a benchmark from the small set. The correlation coefficient is 78\%.  The results show that our CNOT reductions are indicative of fidelity savings.}
\label{fig:correl_q5}
\end{figure}

To measure how much improvement our \SQS{} technique can achieve on the current available NISQ machines, we perform our experiments on quantum devices. Since our medium-size and large-size circuits are too deep for the current quantum devices to generate meaningful results, we choose small-size circuits for this evaluation. We run the small-size benchmarks on IBM's Athens device~\cite{IBMQX}. We use total variation distance, $d_{TV}$, to compare measurement samples of the real device with those of the ideal simulation. Total variation distance is commonly used as a metric for QC experiments \cite{buadescu2019quantum,lund2017quantum,arute2019quantum,ding2020square}. Lower $d_{TV}$ means the samples of the real device are closer to the ideal distribution. Figure~\ref{fig:tvd_q5} shows the results on IBM's Athens, a 5-qubit device; each data point is obtained from 8192 shots of the circuit execution. We observe that \SQS{} achieves lower total variation distance for all benchmarks compared with the baseline optimization, and the distance results are correlated to the number of CNOT reduction. Even though the synthesis distance with 4-qubit blocks is larger than with 3-qubit blocks, \SQS-4 can achieve lower $d_{TV}$ due to more CNOT reduction. Figure~\ref{fig:correl_q5} shows the correlation between CNOT reduction rate and $d_{TV}$ improvement ($\Delta d_{TV}$) compared with the baseline. The correlation coefficient is 78\%. The results show that our CNOT reductions are  indicative of fidelity savings. The results suggest that our \SQS{} optimization technique is important in the NISQ era, and is applicable to the current quantum devices.

\subsection{Running on Noise Simulation}\label{subsec:sim}
\begin{figure}[!t]
\centering
\includegraphics[width=0.42\textwidth,keepaspectratio]{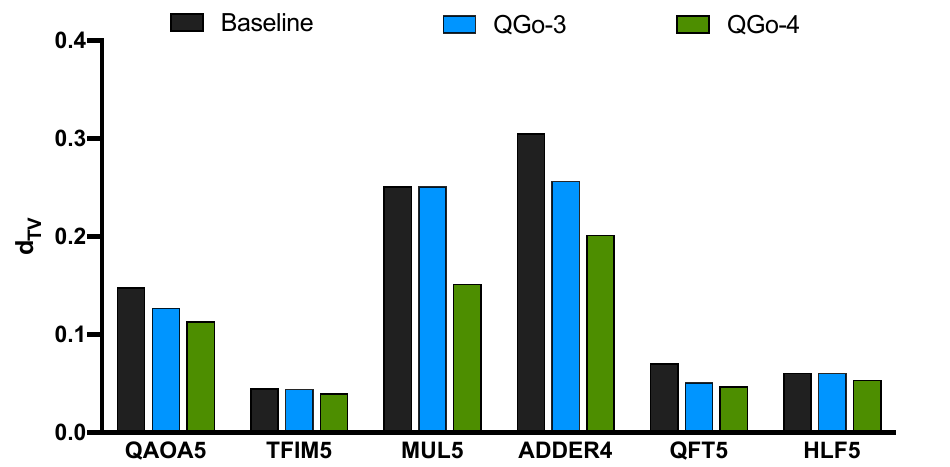}
\caption{Running optimized small circuits on Qiskit noise simulation. (Lower $d_{TV}$ is better.)}
\label{fig:tvd_q5_sim}
\end{figure}

\begin{figure}[!t]
\centering
\includegraphics[width=0.36\textwidth,keepaspectratio]{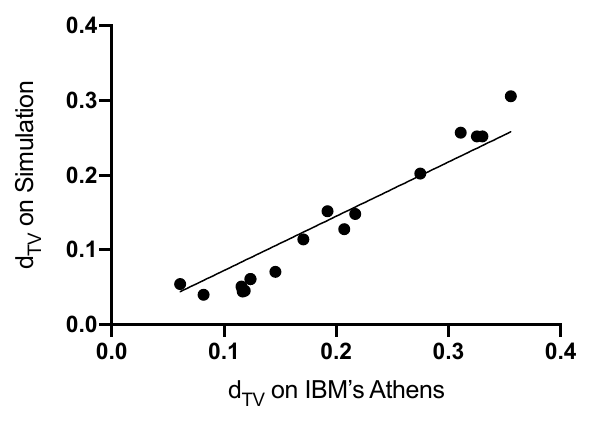}
\caption{Correlation between $d_{TV}$ results on the real device and noise simulation. Each data point is a benchmark from the small set. The correlation coefficient is 98\%.}
\label{fig:correl_q5_sim}
\end{figure}

\begin{figure}[!t]
\centering
\includegraphics[width=0.36\textwidth,keepaspectratio]{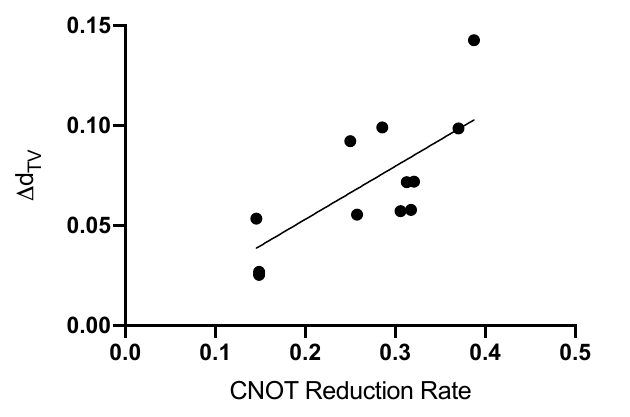}
\caption{Correlation between CNOT reduction rate and $d_{TV}$ improvement measured on noise simulation. Each data point is a benchmark from the medium set. The correlation coefficient is 76\%.}
\label{fig:correl_medium}
\end{figure}

To understand the performance of \SQS{} optimization for medium circuits, we run the experiments on IBM Qiskit Aer noise simulation~\cite{aleksandrowicz2019qiskit} because the medium circuits are too deep to get meaningful results on the current available real devices. To validate the trends shown by our simulations, we run small circuits on both real devices and simulations to see the correlation between them. Figure~\ref{fig:tvd_q5_sim} shows the results from noise simulation. Compared with the results on IBM's Athens (Figure~\ref{fig:tvd_q5}), the $d_{TV}$ improvements are correlated in both experiments. There is a 98\% correlation between our real-device results and our simulation results on small circuits (Figure~\ref{fig:correl_q5_sim}). For medium circuits on simulation, we show the correlation between CNOT reduction rate and $d_{TV}$ improvement compared with the baseline in Figure~\ref{fig:correl_medium}. The correlation coefficient is 76\%.

\begin{figure*}[!t] 
\centering
\includegraphics[width=0.3\textwidth,keepaspectratio]{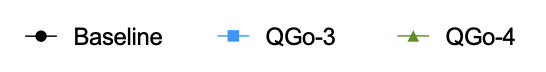}

\subfigure[QAOA10]
{
\includegraphics[width=0.3\textwidth,keepaspectratio]{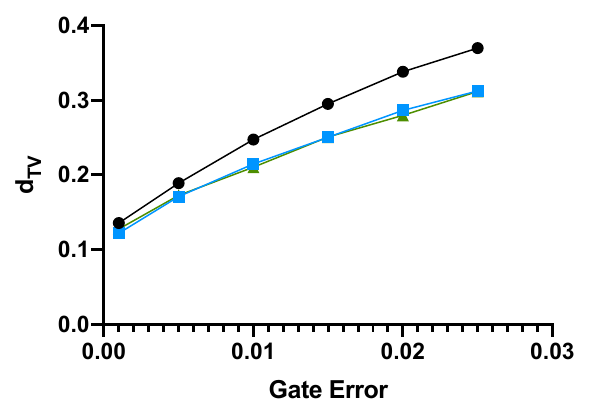}
\label{fig:qaoa_q10_tvd}
}
\quad
\subfigure[TFIM10]
{
\includegraphics[width=0.3\textwidth,keepaspectratio]{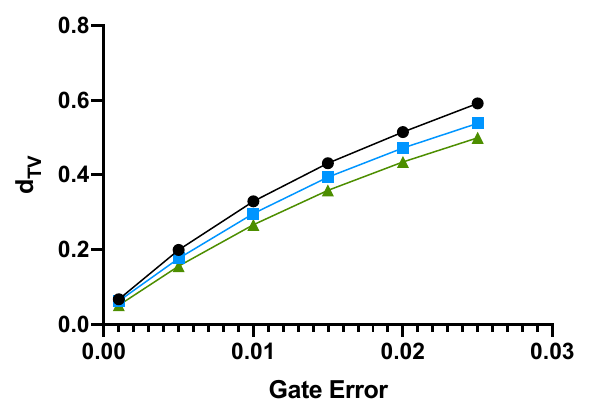}
\label{fig:tfim_q10_tvd}
}
\quad
\subfigure[MUL10]
{
\includegraphics[width=0.3\textwidth,keepaspectratio]{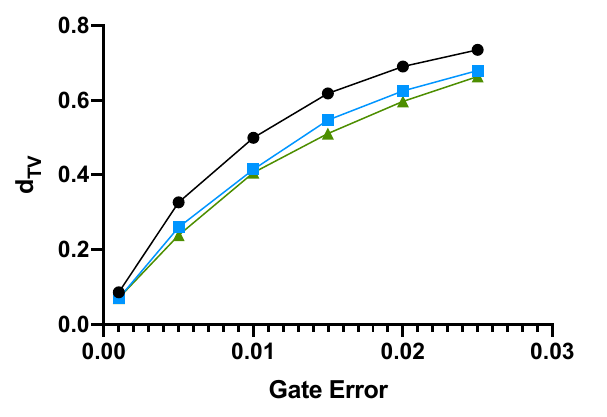}
\label{fig:mul_q10_tvd}
}
\subfigure[ADDER9]
{
\includegraphics[width=0.3\textwidth,keepaspectratio]{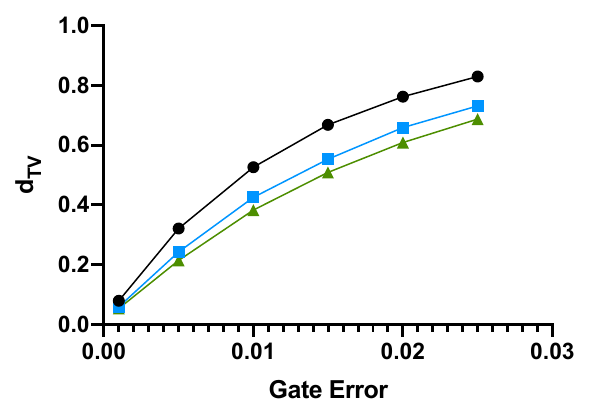}
\label{fig:adder_q9_tvd}
}
\quad
\subfigure[QFT9]
{
\includegraphics[width=0.3\textwidth,keepaspectratio]{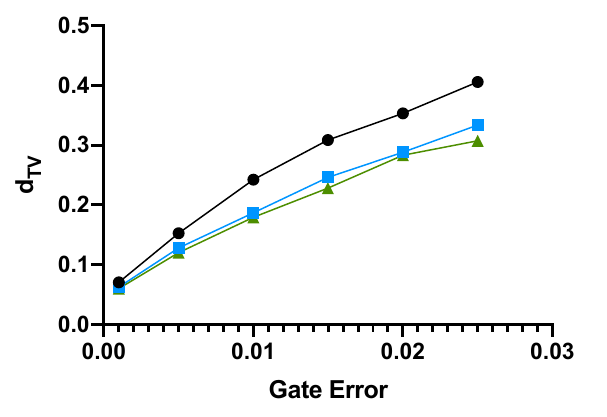}
\label{fig:qft_q9_tvd}
}
\quad
\subfigure[HLF9]
{
\includegraphics[width=0.3\textwidth,keepaspectratio]{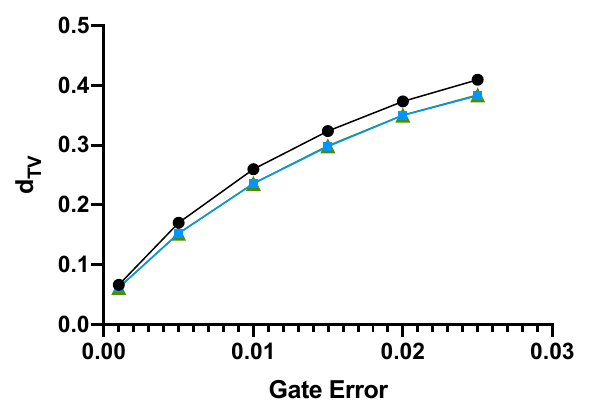}
\label{fig:hlf_q9_tvd}
}

\caption{ Realistic noise simulation results of $d_{TV}$ under different levels of gate error. (Lower $d_{TV}$ is better.)}
\label{fig:tvd}
\end{figure*}

We perform sensitivity analysis by running noise simulations for medium circuits with different gate errors.  Figure~\ref{fig:tvd} shows the noise simulation results. We simulate depolarize\_noise for two-qubit gate noises with the gate error probability from 0.1\% to 2.5\%. We can observe that \SQS{} can reduce more $d_{TV}$ when the gate error is large.

\subsection{Scalability}\label{subsec:model}
\begin{figure}[!t]
\centering
\includegraphics[width=0.42\textwidth,keepaspectratio]{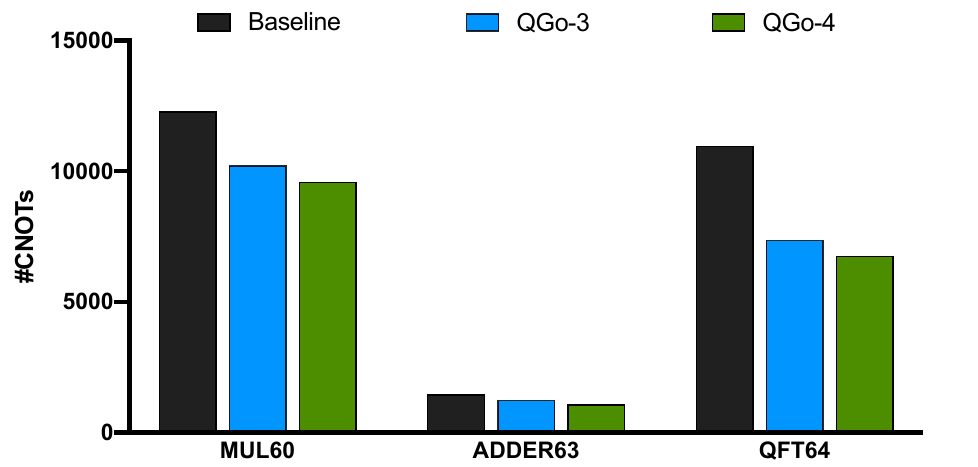}
\caption{The number of CNOTs in the optimized large-scale circuits.}
\label{fig:num_CNOT_large}
\end{figure}

\begin{figure}[!t]
\centering
\includegraphics[width=0.42\textwidth,keepaspectratio]{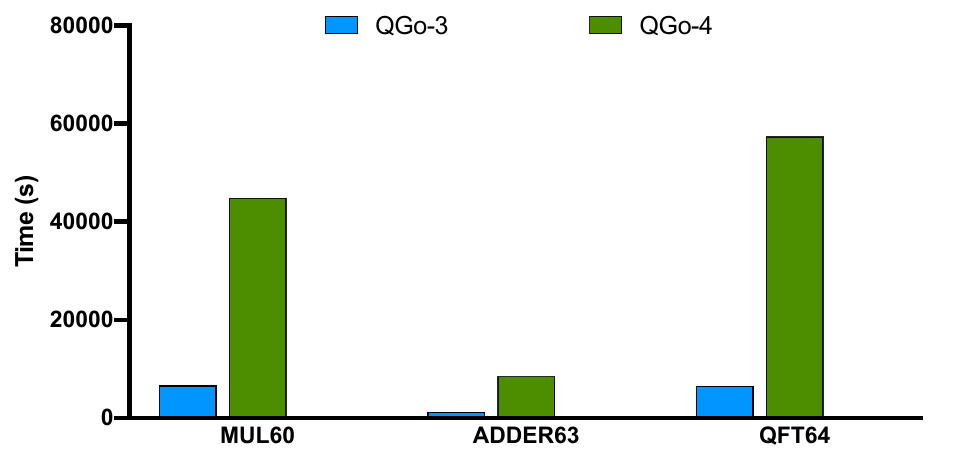}
\caption{Compilation time of large-scale circuit optimization.}
\label{fig:time_large}
\end{figure}

\begin{figure}[!t]
\centering
\includegraphics[width=0.42\textwidth,keepaspectratio]{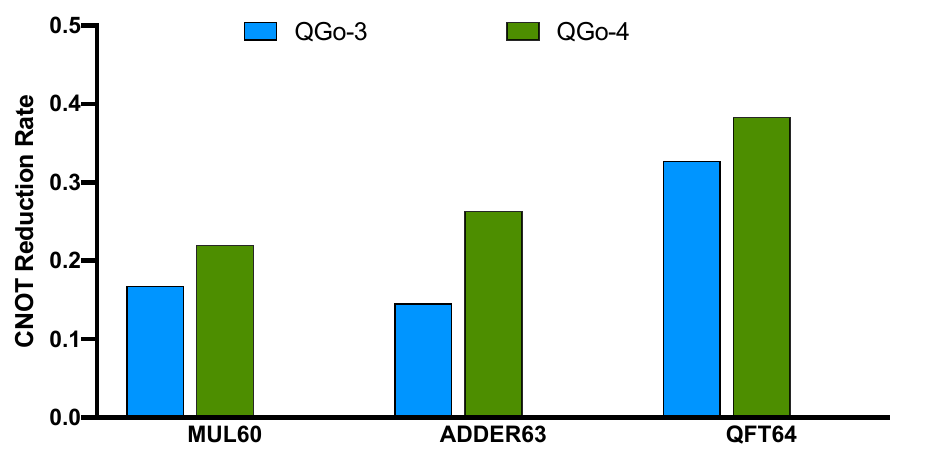}
\caption{CNOT reduction rate of large-scale circuits.}
\label{fig:reduction_large}
\end{figure}

To demonstrate the scalability of our \SQS{} technique, we optimize the large-scale (60+ qubits) benchmarks using \SQS. We map the large-scale circuits on a 2D ($8\times8$) lattice of physical qubits. Figure~\ref{fig:num_CNOT_large} shows the total number of CNOTs optimized by the baseline compiler and \SQS, and Figure~\ref{fig:reduction_large} shows the CNOT reduction rate.  With large-size circuits, \SQS-3 and \SQS-4 achieves 22.8\% and 28.9\% CNOT reduction on average. \SQS-4 performs better than \SQS-3 in terms of CNOT reduction, but it takes longer time to complete the optimization. Figure~\ref{fig:time_large} shows the compilation time of large-scale circuit optimization. \SQS-3{} can complete the optimization within a few minutes, and \SQS-4 can finish the optimization process within a few hours. 

\begin{figure*}[!t] 
\centering
\includegraphics[width=0.3\textwidth,keepaspectratio]{./syn_fig/tvd_legends}

\subfigure[MUL60]
{
\includegraphics[width=0.3\textwidth,keepaspectratio]{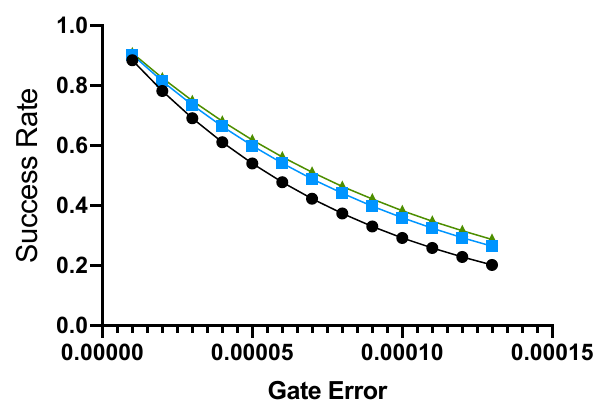}
\label{fig:mul_fid}
}
\quad
\subfigure[ADDER63]
{
\includegraphics[width=0.3\textwidth,keepaspectratio]{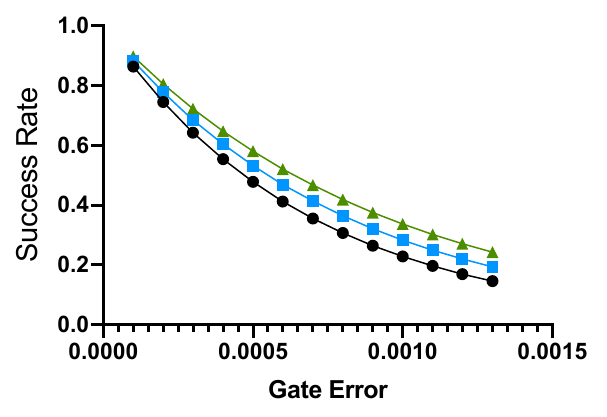}
\label{fig:adder_fid}
}
\quad
\subfigure[QFT64]
{
\includegraphics[width=0.3\textwidth,keepaspectratio]{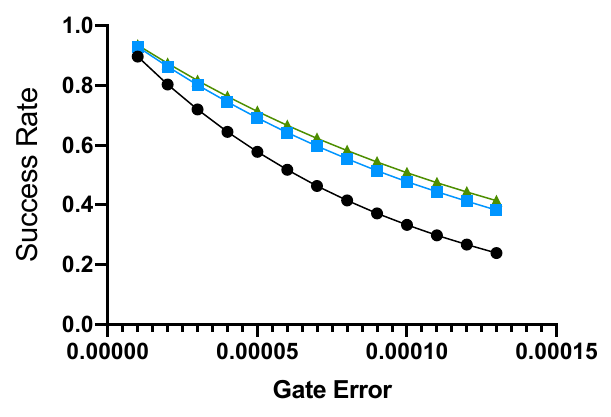}
\label{fig:qft_fid}
}
\caption{Circuit success rate under different levels of gate errors. (Higher circuit success rate is better.)}
\label{fig:fidelity}
\end{figure*}

Since the scale is too large to perform noise simulation, we use an analytical model to estimate the success rate of each circuit. The success rates in our evaluation are computed by a worst-case analysis using gate success rates. Multiplying the gate success rates, we can obtain the estimated success rate for the whole circuit. Figure~\ref{fig:fidelity} shows the results under different gate error models. Since our \SQS-4 has the lowest CNOT count, it is projected to achieve the highest success rates for all benchmarks. The success rate improvement is greater when there is a larger gate error.

\subsection{Discussion}
The results show the general applicability of our approach. Having access to 3-qubit block synthesis already enables good optimization results on large circuits. In general, larger block size can achieve more CNOT reduction. Running synthesis with 5-qubit blocks is possible, but the solving time is much longer.

\SQS{} allows composability with any mapper available for a given platform and we have experimented with  \tket{} compiler. Our preliminary sensitivity analysis of circuit quality to mapping quality indicates there is a direct correlation between \SQS{} efficacy and mapping quality. The best quality mappings have highly interconnected components. In our conjecture, the better mapper provides the higher  opportunity of forming large blocks, thus motivating improvements in both mapping and synthesis. Our study offers insights for future compiler design.  


%% file: T_Extension.tex
\section{Future Work}\label{sec:ext}
Even though \SQS{} already achieves successful results in optimizing different sizes of circuits, we discuss a few directions that would further extend the line of this research and improve the development of quantum computing.

\subsection{Approximate Synthesis}
The objective of optimization using synthesis is to reduce the CNOT count.  In some cases, it may be desirable to make approximations to reduce the number of CNOTs at the expense of how closely the final circuit approximates the desired unitary.  An intuitive way to achieve this is to relax a synthesis threshold.  The synthesis tool that we used is capable of doing this, but even when loosening the threshold as far as $10^-4$, we found no reduction.  The limiting factor on approximation in this case is that the heuristic, process infidelity, is not flexible in a way needed to find good approximations.  In order to effectively trade accuracy for CNOT count, a threshold-controllable synthesis tool is necessary. Integrating a different synthesis tool or heuristic that is designed for approximate synthesis may further improve the optimization. 

\subsection{Pulse-Level Optimization}
Since the device-level control of a quantum computer is operated via analog pulses, recent pulse optimization studies aim to generate shorter pulses~\cite{glaser2015training,shi2019optimized,gokhale2019partial,gokhale2020optimized,DBLP:conf/isca/ChengDQ20}. Pulse optimization can also be integrated into our \SQS{} technique. The current synthesis output is a sequence of quantum gates. To integrate pulse optimization into this work, we can synthesize the block into a sequence of pulses.

\subsection{\SQS{} in Parallel}
In this work, we perform the quantum synthesis in serial. However, since we partition a circuit into a sequence of circuit blocks, each block is independent for the synthesis process. As a result, the synthesis of blocks can be executed in parallel. Running the entire \SQS{} on  a supercomputing system can reduce the overall compilation time significantly.

%% file: T_RelatedWork.tex
\section{Related Work}\label{sec:related}

Several studies of circuit optimization have been carried out. Most of the existing techniques focus on optimizing the qubit mapping and swap insertion to reduce circuit depth. One common approach is to describe the problem in a mathematical form, such as integer linear programming, and then find the optimal solutions by using solvers~\cite{bahreini2015minlp,wille2014optimal,lye2015determining,maslov2008quantum,grosse2009exact}. This approach only works for small circuits since the time scaling is exponential. Another approach is to find the optimal solutions by using dynamic programming ~\cite{siraichi2018collange,itoko2020optimization}. However, since the solving time grows exponentially, this method only works for a handful of qubits. Recent studies propose using heuristic search algorithms to find good solutions to avoid long execution times~\cite{itoko2020optimization, li2019tackling,zulehner2018efficient,alfailakawi2014lnn,saeedi2011synthesis,wille2016look,kole2016heuristic,bhattacharjee2018novel}. However, these approaches keep the original CNOT count and only reduce the additional swap count. Our synthesis approach can reduce both swap count and CNOT count used in the circuits. 

Previous studies have applied synthesis technique to optimize some specific circuits such as classical reversible circuits~\cite{bandyopadhyay2020post,grosse2009exact,wille2019towards,alwardi2018synthesis} or Clifford+T circuits ~\cite{niemann2020advanced,niemann2018improved}. In this work, our approach is designed for general circuits. However, since we can easily change the core synthesis tool, these synthesis approaches can be integrated in our compiler framework to improve the optimization for these specific circuits.

Our approach relies on synthesis techniques that are able to produce extremely short circuits.  Otherwise, it is unlikely that we will be able to see an improvement when resynthesizing sub-circuits.  The KAK decomposition could be used for resynthesizing 2-qubit blocks \cite{tucci2005introduction}.  We have seen improvement when using larger block sizes, so we use the search-based technique found in \cite{davis2019heuristics}, which can handle circuits as large as 4 qubits.  For scaling further, we are considering the approach found in \cite{younis2020qfast}, which produces slightly longer circuits, but offers a better scaling runtime when compared to the search-based approach.  


Recent studies such as \cite{bocharov2015efficient} propose the removal of the constraint of unitary operations by adding ancilla qubits.  Additionally, \cite{camps2020approximate} uses ancillas and an approximation technique to produce very short circuits.  To achieve greater CNOT reduction, integrating ancillas and approximate synthesis into our \SQS{} is a promising research direction.

%% file: T_Conclusion.tex
\section{Conclusion}\label{sec:conclusion}
In the NISQ era, since two-qubit gates are much noisier than single-qubit gates, it is essential to minimize their count. Synthesis is a powerful tool for circuit optimization to produce shorter circuits to improve the overall circuit fidelity. However, synthesis is only applicable for small circuits. In this work, we present an automated compilation framework, \SQS{}. It partitions the circuit into blocks, and re-generates each optimized block by using synthesis, and re-composes the circuit by stitching all the blocks together. Our approach to circuit optimization offers a role for quantum synthesis algorithms in large-scale quantum computing scenarios. We  evaluate  fidelity  improvements  using 3 metrics for 3 scaling regimes: small-size circuits using fidelity on real devices, medium-size circuits using fidelity using simulations with noise, and large-size circuits using CNOT reduction measured statically by our compiler. The results show that our technique has practical value on current devices and is reliable in the NISQ era. We also discuss using approximate synthesis to further trade for circuit depth and pulse-level optimization. Our study of circuit optimization using synthesis offers insights for future compiler design.

%% file: T_Ack.tex
\section*{Acknowledgments}
This work is funded in part by EPiQC, an NSF Expedition in Computing, under grants CCF-1730082/1730449; in part by STAQ under grant NSF Phy-1818914; in part by NSF Grant No. 2110860; in part by the US Department of Energy Office of Advanced Scientific Computing Research, Accelerated Research for Quantum Computing Program; and in part by NSF OMA-2016136 and in part based upon work supported by the 
U.S. Department of Energy, Office of Science, National Quantum Information Science Research Centers. This work was supported by the U.S. Department of Energy (DOE) under Contract No. DE-AC02-05CH11231, through the Office of Advanced Scientific Computing Research Accelerated Research for Quantum Computing Program.

FTC is Chief Scientist at Super.tech and an advisor to Quantum Circuits, Inc.

%% file: main.bbl
\begin{thebibliography}{10}
\providecommand{\url}[1]{#1}
\csname url@samestyle\endcsname
\providecommand{\newblock}{\relax}
\providecommand{\bibinfo}[2]{#2}
\providecommand{\BIBentrySTDinterwordspacing}{\spaceskip=0pt\relax}
\providecommand{\BIBentryALTinterwordstretchfactor}{4}
\providecommand{\BIBentryALTinterwordspacing}{\spaceskip=\fontdimen2\font plus
\BIBentryALTinterwordstretchfactor\fontdimen3\font minus
  \fontdimen4\font\relax}
\providecommand{\BIBforeignlanguage}[2]{{%
\expandafter\ifx\csname l@#1\endcsname\relax
\typeout{** WARNING: IEEEtranS.bst: No hyphenation pattern has been}%
\typeout{** loaded for the language `#1'. Using the pattern for}%
\typeout{** the default language instead.}%
\else
\language=\csname l@#1\endcsname
\fi
#2}}
\providecommand{\BIBdecl}{\relax}
\BIBdecl

\bibitem{al2015quantum}
O.~Al-Ta'ani, ``Quantum circuit synthesis using solovay-kitaev algorithm and
  optimization techniques,'' Ph.D. dissertation, Kansas State University, 2015.

\bibitem{aleksandrowicz2019qiskit}
G.~Aleksandrowicz, T.~Alexander, P.~Barkoutsos, L.~Bello, Y.~Ben-Haim,
  D.~Bucher, F.~Cabrera-Hern{\'a}ndez, J.~Carballo-Franquis, A.~Chen, C.~Chen
  \emph{et~al.}, ``Qiskit: An open-source framework for quantum computing,''
  \emph{Accessed on: Mar}, vol.~16, 2019.

\bibitem{alfailakawi2014lnn}
M.~AlFailakawi, I.~Ahmad, and S.~Hamdan, ``Lnn reversible circuit realization
  using fast harmony search based heuristic,'' in \emph{Asia-Pacific Conference
  on Computer Science and Electrical Engineering}, 2014.

\bibitem{alwardi2018synthesis}
Z.~Alwardi, R.~Wille, and R.~Drechsler, ``Synthesis of reversible circuits
  using conventional hardware description languages,'' in \emph{2018 IEEE 48th
  International Symposium on Multiple-Valued Logic (ISMVL)}.\hskip 1em plus
  0.5em minus 0.4em\relax IEEE, 2018, pp. 97--102.

\bibitem{amy2013meet}
M.~Amy, D.~Maslov, M.~Mosca, and M.~Roetteler, ``A meet-in-the-middle algorithm
  for fast synthesis of depth-optimal quantum circuits,'' \emph{IEEE
  Transactions on Computer-Aided Design of Integrated Circuits and Systems},
  vol.~32, no.~6, pp. 818--830, 2013.

\bibitem{arute2019quantum}
F.~Arute, K.~Arya, R.~Babbush, D.~Bacon, J.~C. Bardin, R.~Barends, R.~Biswas,
  S.~Boixo, F.~G. Brandao, D.~A. Buell \emph{et~al.}, ``Quantum supremacy using
  a programmable superconducting processor,'' \emph{Nature}, vol. 574, no.
  7779, pp. 505--510, 2019.

\bibitem{buadescu2019quantum}
C.~B{\u{a}}descu, R.~O'Donnell, and J.~Wright, ``Quantum state certification,''
  in \emph{Proceedings of the 51st Annual ACM SIGACT Symposium on Theory of
  Computing}, 2019, pp. 503--514.

\bibitem{bahreini2015minlp}
T.~Bahreini and N.~Mohammadzadeh, ``An {MINLP} model for scheduling and
  placement of quantum circuits with a heuristic solution approach,'' \emph{ACM
  Journal on Emerging Technologies in Computing Systems (JETC)}, vol.~12,
  no.~3, pp. 1--20, 2015.

\bibitem{bandyopadhyay2020post}
C.~Bandyopadhyay, R.~Wille, R.~Drechsler, and H.~Rahaman, ``Post
  synthesis-optimization of reversible circuit using template matching,'' in
  \emph{2020 24th International Symposium on VLSI Design and Test
  (VDAT)}.\hskip 1em plus 0.5em minus 0.4em\relax IEEE, 2020, pp. 1--4.

\bibitem{bhattacharjee2018novel}
A.~Bhattacharjee, C.~Bandyopadhyay, R.~Wille, R.~Drechsler, and H.~Rahaman, ``A
  novel approach for nearest neighbor realization of {2D} quantum circuits,''
  in \emph{2018 IEEE Computer Society Annual Symposium on VLSI (ISVLSI)}.\hskip
  1em plus 0.5em minus 0.4em\relax IEEE, 2018, pp. 305--310.

\bibitem{biamonte2017quantum}
J.~Biamonte, P.~Wittek, N.~Pancotti, P.~Rebentrost, N.~Wiebe, and S.~Lloyd,
  ``Quantum machine learning,'' \emph{Nature}, vol. 549, no. 7671, p. 195,
  2017.

\bibitem{bocharov2015efficient}
A.~Bocharov, M.~Roetteler, and K.~M. Svore, ``Efficient synthesis of universal
  repeat-until-success quantum circuits,'' \emph{Physical review letters}, vol.
  114, no.~8, p. 080502, 2015.

\bibitem{bravyi2018quantum}
S.~Bravyi, D.~Gosset, and R.~K{\"o}nig, ``Quantum advantage with shallow
  circuits,'' \emph{Science}, vol. 362, no. 6412, pp. 308--311, 2018.

\bibitem{camps2020approximate}
D.~Camps and R.~Van~Beeumen, ``Approximate quantum circuit synthesis using
  block encodings,'' \emph{Physical Review A}, vol. 102, no.~5, p. 052411,
  2020.

\bibitem{chakrabarti2011linear}
A.~Chakrabarti, S.~Sur-Kolay, and A.~Chaudhury, ``Linear nearest neighbor
  synthesis of reversible circuits by graph partitioning,'' \emph{arXiv
  preprint arXiv:1112.0564}, 2011.

\bibitem{DBLP:conf/isca/ChengDQ20}
\BIBentryALTinterwordspacing
J.~Cheng, H.~Deng, and X.~Qian, ``Accqoc: Accelerating quantum optimal control
  based pulse generation,'' in \emph{47th {ACM/IEEE} Annual International
  Symposium on Computer Architecture, {ISCA} 2020, Valencia, Spain, May 30 -
  June 3, 2020}.\hskip 1em plus 0.5em minus 0.4em\relax {IEEE}, 2020, pp.
  543--555. [Online]. Available:
  \url{https://doi.org/10.1109/ISCA45697.2020.00052}
\BIBentrySTDinterwordspacing

\bibitem{cleve1998quantum}
R.~Cleve, A.~Ekert, C.~Macchiavello, and M.~Mosca, ``Quantum algorithms
  revisited,'' \emph{Proceedings of the Royal Society of London. Series A:
  Mathematical, Physical and Engineering Sciences}, vol. 454, no. 1969, pp.
  339--354, 1998.

\bibitem{cowtan2019phase}
A.~Cowtan, S.~Dilkes, R.~Duncan, W.~Simmons, and S.~Sivarajah, ``Phase gadget
  synthesis for shallow circuits,'' \emph{arXiv preprint arXiv:1906.01734},
  2019.

\bibitem{cross2018ibm}
A.~Cross, ``The {IBM Q experience and QISKit} open-source quantum computing
  software,'' \emph{Bulletin of the American Physical Society}, vol.~63, 2018.

\bibitem{cuccaro2004new}
S.~A. Cuccaro, T.~G. Draper, S.~A. Kutin, and D.~P. Moulton, ``A new quantum
  ripple-carry addition circuit,'' \emph{arXiv preprint quant-ph/0410184},
  2004.

\bibitem{davis2019heuristics}
M.~G. {Davis}, E.~{Smith}, A.~{Tudor}, K.~{Sen}, I.~{Siddiqi}, and C.~{Iancu},
  ``Towards optimal topology aware quantum circuit synthesis,'' in \emph{2020
  IEEE International Conference on Quantum Computing and Engineering (QCE)},
  2020, pp. 223--234.

\bibitem{dawson2005solovay}
C.~M. Dawson and M.~A. Nielsen, ``The solovay-kitaev algorithm,'' \emph{arXiv
  preprint quant-ph/0505030}, 2005.

\bibitem{de2016block}
A.~De~Vos and S.~De~Baerdemacker, ``Block-z x z synthesis of an arbitrary
  quantum circuit,'' \emph{Physical Review A}, vol.~94, no.~5, p. 052317, 2016.

\bibitem{di2016parallelizing}
O.~Di~Matteo and M.~Mosca, ``Parallelizing quantum circuit synthesis,''
  \emph{Quantum Science and Technology}, vol.~1, no.~1, p. 015003, 2016.

\bibitem{ding2020square}
Y.~Ding, X.-C. Wu, A.~Holmes, A.~Wiseth, D.~Franklin, M.~Martonosi, and F.~T.
  Chong, ``Square: Strategic quantum ancilla reuse for modular quantum programs
  via cost-effective uncomputation,'' in \emph{ACM/IEEE 47th Annual
  International Symposium on Computer Architecture (ISCA)}, 2020.

\bibitem{divincenzo1995two}
D.~P. DiVincenzo, ``Two-bit gates are universal for quantum computation,''
  \emph{Physical Review A}, vol.~51, no.~2, p. 1015, 1995.

\bibitem{farhi2014quantum}
E.~Farhi, J.~Goldstone, and S.~Gutmann, ``A quantum approximate optimization
  algorithm,'' \emph{arXiv preprint arXiv:1411.4028}, 2014.

\bibitem{gidneycirq}
C.~Gidney and D.~Bacon, ``The cirq developers, quantumlib/-cirq: A python
  framework for creating, editing, and invoking noisy intermediate scale
  quantum (nisq) circuits.''

\bibitem{giles2013exact}
B.~Giles and P.~Selinger, ``Exact synthesis of multiqubit clifford+ t
  circuits,'' \emph{Physical Review A}, vol.~87, no.~3, p. 032332, 2013.

\bibitem{glaser2015training}
S.~J. Glaser, U.~Boscain, T.~Calarco, C.~P. Koch, W.~K{\"o}ckenberger,
  R.~Kosloff, I.~Kuprov, B.~Luy, S.~Schirmer, T.~Schulte-Herbr{\"u}ggen
  \emph{et~al.}, ``Training schr{\"o}dinger’s cat: quantum optimal control,''
  \emph{The European Physical Journal D}, vol.~69, no.~12, pp. 1--24, 2015.

\bibitem{gokhale2019partial}
P.~Gokhale, Y.~Ding, T.~Propson, C.~Winkler, N.~Leung, Y.~Shi, D.~I. Schuster,
  H.~Hoffmann, and F.~T. Chong, ``Partial compilation of variational algorithms
  for noisy intermediate-scale quantum machines,'' in \emph{Proceedings of the
  52nd Annual IEEE/ACM International Symposium on Microarchitecture}, 2019, pp.
  266--278.

\bibitem{gokhale2020optimized}
P.~Gokhale, A.~Javadi-Abhari, N.~Earnest, Y.~Shi, and F.~T. Chong, ``Optimized
  quantum compilation for near-term algorithms with openpulse,'' \emph{arXiv
  preprint arXiv:2004.11205}, 2020.

\bibitem{Google}
``{A} {P}review of {B}ristlecone, {G}oogle's {N}ew {Q}uantum {P}rocessor,''
  \url{https://ai.googleblog.com/2018/03/a-preview-of-bristlecone-googles-new.html},
  accessed: 2020-10-09.

\bibitem{grosse2009exact}
D.~Gro{\ss}e, R.~Wille, G.~W. Dueck, and R.~Drechsler, ``Exact multiple-control
  toffoli network synthesis with sat techniques,'' \emph{IEEE Transactions on
  Computer-Aided Design of Integrated Circuits and Systems}, vol.~28, no.~5,
  pp. 703--715, 2009.

\bibitem{heyl2013dynamical}
M.~Heyl, A.~Polkovnikov, and S.~Kehrein, ``Dynamical quantum phase transitions
  in the transverse-field ising model,'' \emph{Physical review letters}, vol.
  110, no.~13, p. 135704, 2013.

\bibitem{IBM}
``{IBM} {A}nnounces {A}dvances to {IBM} {Q}uantum {S}ystems and {E}cosystem,''
  \url{https://www-03.ibm.com/press/us/en/pressrelease/53374.wss}, accessed:
  2020-10-09.

\bibitem{IBMQX}
``{IBM} {Q}uantum {E}xperience,'' \url{https://quantum-computing.ibm.com/},
  accessed: 2020-11-15.

\bibitem{itoko2020optimization}
T.~Itoko, R.~Raymond, T.~Imamichi, and A.~Matsuo, ``Optimization of quantum
  circuit mapping using gate transformation and commutation,''
  \emph{Integration}, vol.~70, pp. 43--50, 2020.

\bibitem{javadiabhari2015scaffcc}
A.~JavadiAbhari, S.~Patil, D.~Kudrow, J.~Heckey, A.~Lvov, F.~T. Chong, and
  M.~Martonosi, ``{ScaffCC}: Scalable compilation and analysis of quantum
  programs,'' \emph{Parallel Computing}, vol.~45, pp. 2--17, 2015.

\bibitem{jozsa2001quantum}
R.~Jozsa, ``Quantum factoring, discrete logarithms, and the hidden subgroup
  problem,'' \emph{Computing in science \& engineering}, vol.~3, no.~2, p.~34,
  2001.

\bibitem{kandala2017hardware}
A.~Kandala, A.~Mezzacapo, K.~Temme, M.~Takita, M.~Brink, J.~M. Chow, and J.~M.
  Gambetta, ``Hardware-efficient variational quantum eigensolver for small
  molecules and quantum magnets,'' \emph{Nature}, vol. 549, no. 7671, p. 242,
  2017.

\bibitem{kliuchnikov2012fast}
V.~Kliuchnikov, D.~Maslov, and M.~Mosca, ``Fast and efficient exact synthesis
  of single qubit unitaries generated by clifford and t gates,'' \emph{arXiv
  preprint arXiv:1206.5236}, 2012.

\bibitem{kole2016heuristic}
A.~Kole, K.~Datta, and I.~Sengupta, ``A heuristic for linear nearest neighbor
  realization of quantum circuits by {SWAP} gate insertion using {N}-gate
  lookahead,'' \emph{IEEE Journal on Emerging and Selected Topics in Circuits
  and Systems}, vol.~6, no.~1, pp. 62--72, 2016.

\bibitem{li2019tackling}
G.~Li, Y.~Ding, and Y.~Xie, ``Tackling the qubit mapping problem for nisq-era
  quantum devices,'' in \emph{Proceedings of the Twenty-Fourth International
  Conference on Architectural Support for Programming Languages and Operating
  Systems}, 2019, pp. 1001--1014.

\bibitem{lund2017quantum}
A.~P. Lund, M.~J. Bremner, and T.~C. Ralph, ``Quantum sampling problems,
  bosonsampling and quantum supremacy,'' \emph{npj Quantum Information},
  vol.~3, no.~1, pp. 1--8, 2017.

\bibitem{lye2015determining}
A.~Lye, R.~Wille, and R.~Drechsler, ``Determining the minimal number of swap
  gates for multi-dimensional nearest neighbor quantum circuits,'' in \emph{The
  20th Asia and South Pacific Design Automation Conference}.\hskip 1em plus
  0.5em minus 0.4em\relax IEEE, 2015, pp. 178--183.

\bibitem{martinez2016compiling}
E.~A. Martinez, T.~Monz, D.~Nigg, P.~Schindler, and R.~Blatt, ``Compiling
  quantum algorithms for architectures with multi-qubit gates,'' \emph{New
  Journal of Physics}, vol.~18, no.~6, p. 063029, 2016.

\bibitem{maslov2008quantum}
D.~Maslov, S.~M. Falconer, and M.~Mosca, ``Quantum circuit placement,''
  \emph{IEEE Transactions on Computer-Aided Design of Integrated Circuits and
  Systems}, vol.~27, no.~4, pp. 752--763, 2008.

\bibitem{moll2018quantum}
N.~Moll, P.~Barkoutsos, L.~S. Bishop, J.~M. Chow, A.~Cross, D.~J. Egger,
  S.~Filipp, A.~Fuhrer, J.~M. Gambetta, M.~Ganzhorn \emph{et~al.}, ``Quantum
  optimization using variational algorithms on near-term quantum devices,''
  \emph{Quantum Science and Technology}, vol.~3, no.~3, p. 030503, 2018.

\bibitem{nagy2006implementation}
A.~B. Nagy, ``On an implementation of the solovay-kitaev algorithm,''
  \emph{arXiv preprint quant-ph/0606077}, 2006.

\bibitem{nielsen2002quantum}
M.~A. Nielsen and I.~Chuang, ``Quantum computation and quantum information,''
  2002.

\bibitem{niemann2018improved}
P.~Niemann, R.~Wille, and R.~Drechsler, ``Improved synthesis of clifford+ t
  quantum functionality,'' in \emph{2018 Design, Automation \& Test in Europe
  Conference \& Exhibition (DATE)}.\hskip 1em plus 0.5em minus 0.4em\relax
  IEEE, 2018, pp. 597--600.

\bibitem{niemann2020advanced}
P.~Niemann, R.~Wille, and R.~Drechsler, ``Advanced exact synthesis of clifford+
  t circuits,'' \emph{Quantum Information Processing}, vol.~19, no.~9, pp.
  1--23, 2020.

\bibitem{nishio2019extracting}
S.~Nishio, Y.~Pan, T.~Satoh, H.~Amano, and R.~Van~Meter, ``Extracting success
  from ibm's 20-qubit machines using error-aware compilation,'' \emph{arXiv
  preprint arXiv:1903.10963}, 2019.

\bibitem{peruzzo2014variational}
A.~Peruzzo, J.~McClean, P.~Shadbolt, M.-H. Yung, X.-Q. Zhou, P.~J. Love,
  A.~Aspuru-Guzik, and J.~L. O’brien, ``A variational eigenvalue solver on a
  photonic quantum processor,'' \emph{Nature Communications}, vol.~5, p. 4213,
  2014.

\bibitem{preskill2018quantum}
J.~Preskill, ``Quantum computing in the {NISQ} era and beyond,''
  \emph{Quantum}, vol.~2, p.~79, 2018.

\bibitem{saeedi2011synthesis}
M.~Saeedi, R.~Wille, and R.~Drechsler, ``Synthesis of quantum circuits for
  linear nearest neighbor architectures,'' \emph{Quantum Information
  Processing}, vol.~10, no.~3, pp. 355--377, 2011.

\bibitem{shende2006synthesis}
V.~V. Shende, S.~S. Bullock, and I.~L. Markov, ``Synthesis of quantum-logic
  circuits,'' \emph{IEEE Transactions on Computer-Aided Design of Integrated
  Circuits and Systems}, vol.~25, no.~6, pp. 1000--1010, 2006.

\bibitem{shi2019optimized}
Y.~Shi, N.~Leung, P.~Gokhale, Z.~Rossi, D.~I. Schuster, H.~Hoffmann, and F.~T.
  Chong, ``Optimized compilation of aggregated instructions for realistic
  quantum computers,'' in \emph{Proceedings of the Twenty-Fourth International
  Conference on Architectural Support for Programming Languages and Operating
  Systems}, 2019, pp. 1031--1044.

\bibitem{shin2018phonon}
D.~Shin, H.~H{\"u}bener, U.~De~Giovannini, H.~Jin, A.~Rubio, and N.~Park,
  ``Phonon-driven spin-floquet magneto-valleytronics in mos 2,'' \emph{Nature
  communications}, vol.~9, no.~1, pp. 1--8, 2018.

\bibitem{shor1999polynomial}
P.~W. Shor, ``Polynomial-time algorithms for prime factorization and discrete
  logarithms on a quantum computer,'' \emph{SIAM review}, vol.~41, no.~2, pp.
  303--332, 1999.

\bibitem{siraichi2018qubit}
M.~Y. Siraichi, V.~F.~d. Santos, S.~Collange, and F.~M.~Q. Pereira, ``Qubit
  allocation,'' in \emph{Proceedings of the 2018 International Symposium on
  Code Generation and Optimization}, 2018, pp. 113--125.

\bibitem{siraichi2018collange}
M.~Y. Siraichi, V.~F.~d. Santos, S.~Collange, and F.~M.~Q. Pereira, ``Qubit
  allocation,'' in \emph{Proceedings of the 2018 International Symposium on
  Code Generation and Optimization}, 2018, pp. 113--125.

\bibitem{sivarajah2020t}
S.~Sivarajah, S.~Dilkes, A.~Cowtan, W.~Simmons, A.~Edgington, and R.~Duncan,
  ``t| ket>: A retargetable compiler for nisq devices,'' \emph{Quantum Science
  and Technology}, 2020.

\bibitem{tucci2005introduction}
R.~R. Tucci, ``An introduction to cartan's kak decomposition for qc
  programmers,'' \emph{arXiv preprint quant-ph/0507171}, 2005.

\bibitem{wille2019towards}
R.~Wille, M.~Haghparast, S.~Adarsh, and M.~Tanmay, ``Towards hdl-based
  synthesis of reversible circuits with no additional lines,'' in \emph{2019
  IEEE/ACM International Conference on Computer-Aided Design (ICCAD)}.\hskip
  1em plus 0.5em minus 0.4em\relax IEEE, 2019, pp. 1--7.

\bibitem{wille2016look}
R.~Wille, O.~Keszocze, M.~Walter, P.~Rohrs, A.~Chattopadhyay, and R.~Drechsler,
  ``Look-ahead schemes for nearest neighbor optimization of {1D and 2D} quantum
  circuits,'' in \emph{2016 21st Asia and South Pacific Design Automation
  Conference (ASP-DAC)}.\hskip 1em plus 0.5em minus 0.4em\relax IEEE, 2016, pp.
  292--297.

\bibitem{wille2014optimal}
R.~Wille, A.~Lye, and R.~Drechsler, ``Optimal {SWAP} gate insertion for nearest
  neighbor quantum circuits,'' in \emph{2014 19th Asia and South Pacific Design
  Automation Conference (ASP-DAC)}.\hskip 1em plus 0.5em minus 0.4em\relax
  IEEE, 2014, pp. 489--494.

\bibitem{wu2020tilt}
X.-C. Wu, D.~M. Debroy, Y.~Ding, J.~M. Baker, Y.~Alexeev, K.~R. Brown, and
  F.~T. Chong, ``Tilt: Achieving higher fidelity on a trapped-ion linear-tape
  quantum computing architecture,'' 2020.

\bibitem{younis2020qfast}
E.~Younis, K.~Sen, K.~Yelick, and C.~Iancu, ``Qfast: Quantum synthesis using a
  hierarchical continuous circuit space,'' \emph{arXiv preprint
  arXiv:2003.04462}, 2020.

\bibitem{zulehner2018efficient}
A.~Zulehner, A.~Paler, and R.~Wille, ``Efficient mapping of quantum circuits to
  the ibm qx architectures,'' in \emph{2018 Design, Automation \& Test in
  Europe Conference \& Exhibition}.\hskip 1em plus 0.5em minus 0.4em\relax
  IEEE, 2018, pp. 1135--1138.

\end{thebibliography}
